\documentclass[structabstract]{aa}  
\usepackage{graphicx}
\usepackage{txfonts}
\usepackage{natbib}
\bibpunct{(}{)}{;}{a}{}{,} 

\begin{document}
   \title{Mid-infrared interferometric monitoring of evolved stars}

   \subtitle{The dust shell around the Mira variable RR~Aql at 13 epochs\thanks{Based on 
observations made with Very Large Telescope Interferometer (VLTI) at the 
Paranal Observatory under program IDs 073.D-0711, 075.D-0097, 077.D-0630, 
and 079.D-0172.}}

   \author{I. Karovicova\inst{1},
          \
          M. Wittkowski\inst{1},
          \
          D. A. Boboltz\inst{2},
          \
	  E. Fossat\inst{3},
          \
          K. Ohnaka\inst{4},
          \and
          M. Scholz\inst{5,6}}

   \institute{European Southern Observatory, Karl-Schwarzschild-Str. 2, 85748 Garching bei M\"unchen, Germany
              \email{ikarovic@eso.org, mwittkow@eso.org}
         \and 
             United States Naval Observatory, 3450 Massachusetts Avenue, NW,Washington, DC 20392-5420, USA
         \and 
	     Laboratoire Univ. d'Astroph. de Nice (LUAN), CNRS UMR 6525, Parc Valrose, 06108 Nice Cedex 02, France
         \and 
             Max-Planck-Institut f\"ur Radioastronomie, Auf dem H\"ugel 69, 53121 Bonn, Germany
         \and
             Zentrum f\"ur Astronomie der Universit\"at Heidelberg (ZAH), Institut f\"ur Theoretische Astrophysik, Albert-Ueberle-Str. 2, 69120 Heidelberg, Germany
         \and 
             Sydney Institute for Astronomy, School of Physics, University of Sydney, Sydney NSW 2006, Australia
             }

   \date{Received .....; accepted ....}

  \abstract
   {}
   {We present a unique multi-epoch infrared interferometric 
    study of the oxygen-rich Mira variable RR~Aql in comparison 
    to radiative transfer models of the dust shell. 
    We investigate 
    flux and visibility spectra at 8 -- 13\,$\mu$m with the 
    aim of better understanding the pulsation mechanism and its 
    connection to the dust condensation sequence and 
    mass-loss process.}
   {We obtained 13 epochs of mid-infrared interferometry with the 
    MIDI instrument at the VLTI between April 2004 and July 2007,
    covering minimum to pre-maximum pulsation phases (0.45--0.85) 
    within 4 cycles. 
    The data are modeled with a radiative transfer model of the dust shell where the central stellar intensity profile is described by a series of dust-free dynamic
model atmospheres based on self-excited pulsation models. 
    We examined two dust species, silicate and Al$_2$O$_3$ grains. 
   We performed model simulations using variations
   in model phase and dust shell parameters
   to investigate the expected variability of our
   mid-infrared photometric 
   and interferometric data. 
}
   {The observed visibility spectra do not show any indication of variations as a function of pulsation phase and cycle. 
The observed photometry spectra may indicate intracycle and cycle-to-cycle variations at the level of 1--2 standard deviations. The photometric and visibility spectra of RR~Aql can be described well by the radiative transfer model of the dust shell that
   uses a dynamic model atmosphere describing the central source. 
    The best-fitting model for our average pulsation phase of $\overline{\Phi_V}=0.64\pm0.15$ includes the dynamic model
   atmosphere M21n ($T_\mathrm{model}=2550$\,K)
   with a photospheric
   angular diameter of $\theta_\mathrm{Phot}=7.6\pm0.6$\,mas,
   and a silicate dust shell with an optical depth of
   $\tau_V=2.8\pm0.8$, an inner radius of
   $R_\mathrm{in}=4.1\pm0.7\,R_\mathrm{Phot}$, and
   a power-law index of the density distribution of $p=2.6\pm0.3$.
   The addition of an Al$_2$O$_3$ dust shell did not improve the
   model fit. However, our model simulations indicate that the presence
   of an inner Al$_2$O$_3$ dust shell with lower optical depth than for the
   silicate dust shell can not be excluded.
   The photospheric angular diameter corresponds to a
   radius of $R_\mathrm{phot}=520^{+230}_{-140}R_\odot$
   and an effective temperature of $T_\mathrm{eff}\sim2420\pm200$\,K.
   Our modeling simulations 
   confirm that significant intracycle and cycle-to-cycle 
   visibility variations are not expected for RR~Aql at mid-infrared 
   wavelengths within our uncertainties.}
   {We conclude that our RR~Aql data can be described by a
   pulsating atmosphere surrounded by a silicate dust shell. The 
   effects of the pulsation on the mid-infrared flux and
   visibility values are expected to be less than about
   25\% and 20\%, respectively, and are too low to be detected
   within our measurement uncertainties. 
   }


   \authorrunning{Karovicova et al.}
   \titlerunning{Mid-infrared interferometric monitoring of the Mira 
    variable RR~Aql}

   \maketitle


\section{Introduction}
Asymptotic giant branch (AGB) stars are low-to-intermediate mass stars 
at the end of their stellar evolution. These stars, including Mira type stars, 
exhibit many complex processes such as shock fronts propagating through 
the stellar atmosphere, large-amplitude pulsation, and molecule and dust 
formation leading to strong mass loss via a dense and dusty outflow 
from an extended stellar atmosphere \citep{Andersen2003}. 
The mass loss rates can reach up to $10^{-4}M_\odot$yr$^{-1}$ 
\citep{Matsuura2009} with expansion velocities of 
5-30 km s$^{-1}$ \citep{Hofner2007}. Dust grains are created 
by condensation from the gas phase in the very cool and dense outer 
part of the pulsating atmosphere. Dust created by 
carbon-rich stars owing to its high opacity absorbs radiation from the 
star, and the wind is believed to be driven by radiation pressure on the 
dust particles that drag gas along via collisions (H{\"o}fner 2008).
 
Despite remarkable progress in theoretical and observational studies, 
it still remains unclear whether the dust in oxygen-rich envelopes 
is opaque enough to drive the observed massive outflows
\citep{Woitke2006,Hofner2008}. Because of the wind, all the matter around 
the core of the star is eventually returned to the interstellar medium, 
and the star evolves toward the planetary nebula phase, leaving the core 
as a white dwarf. Generally many aspects of the physics of AGB stars 
remain poorly understood, such as the detailed atmospheric stratification 
and composition of the stars, or the role of stellar pulsation and its 
connection to the dust formation and massive outflow. Improving our 
understanding of the physical processes leading to the substantial mass 
loss is important, as AGB stars play a crucial role in the chemical 
enrichment and evolution of galaxies by returning gas and dust to the 
interstellar medium (ISM).

Thanks to their large diameters and high luminosities, Mira variables are 
ideal targets for high angular resolution observations. Near-infrared 
interferometric (NIR) techniques provide detailed information regarding 
the conditions near the continuum-forming photosphere such as effective 
temperature, center-to-limb intensity variations, the stellar 
photospheric diameter, and its dependence on wavelength and pulsation 
phase \citep[e.g.][]{Haniff1995,Perrin1999,vanBelle1996,Young2000a,Thompson2002a,Thompson2002b,Quirrenbach1992,Fedele2005,Ohnaka2004,Millan-Gabet2005,Woodruff2008}. New theoretical 
self-excited dynamic model atmospheres of oxygen-rich stars have been 
created and successfully applied \citep{Tej2003,Hofmann1998,Ireland2004a,Ireland2004b,Ireland2008,Wittkowski2008}. 

Observations at mid-infrared wavelengths are well-suited to studying the 
molecular shells and the dust formation zone of evolved stars. Mid-infrared 
interferometry is sensitive to the chemical composition 
and geometry of dust shells, their temperature, inner radii, 
radial distribution, and the mass loss rate \citep[e.g.][]{Danchi1994,Monnier1997,Monnier2000,Weiner2006}. \citet{Lopez1997} 
presented long-term observations at 11\,$\mu$m of o Ceti obtained with 
the Infrared Spatial Interferometer (ISI). The observed visibilities 
change from one epoch to the next and are not consistent 
with simple heating or cooling of the dust with change in luminosity 
as a function of stellar phase, but rather with large temporal variations 
in the density of the dust shell. The data were compared to axially 
symmetric radiative transfer models and suggest inhomogeneities or clumps. 
\citet{Tevousjan2004} has studied the spatial distribution of dust around 
four late-type stars with ISI at 11.15\,$\mu$m, and find that the 
visibility curves change with the pulsation phase of the star. 
The dust grains were modeled as a mixture of silicates and graphite. 
The results suggest that the dust shells appear to be closer to the 
star at minimum pulsation phase, and farther away at maximum phase, which is
also demonstrated by \citet{Wittkowski2007}. 
\citet{Ohnaka2007} reported on temporal visibility variations of the 
carbon-rich Mira variable V Oph with the instruments VINCI and MIDI 
at the VLTI. This temporal variation of the N-band angular sizes is 
largely governed by the variations in the opacity and the 
geometrical extension of the molecular layers (C$_2$H$_2$ and HCN) and the dust shell (amC + SiC).

Additional information to the near- and mid-infrared observations can be 
obtained with Very Long Baseline Array (VLBA) observations, which allow 
one to study the properties of the circumstellar environment of evolved stars 
using the maser radiation emitted by some molecules, most commonly 
SiO, H$_2$O, and OH. By observing the maser radiation at tens to several 
hundred AU from the star, it is possible to determine accurate proper 
motions and parallaxes \citep{Vlemmings2003}, total intensity and 
linear polarizations \citep{Cotton2008}, and the structure of the 
environment of the stars \citep{Diamond1994}. 

This work is part of the ongoing project of concurrent multiwavelength 
observations using optical long baseline interferometry 
(AMBER and MIDI at the VLTI) combined with radio interferometry (VLBA). 
We aim to investigate layers at different depths of the atmosphere 
and circumstellar envelope (CSE) of the stars. Previous results from this project of 
joint VLTI/VLBA observations for the Mira star S Ori can be found 
in \citet{Boboltz2005} and \citet{Wittkowski2007}. 
Here we present results of a long-term VLTI/MIDI monitoring of the 
oxygen-rich Mira variable RR~Aql. In addition to the VLTI/MIDI observation 
presented in this paper, we obtained coordinated multi-epoch VLBA 
observations of SiO and H$_2$O masers towards RR~Aql, which will be 
presented in a subsequent paper.

\begin{figure}
  \includegraphics[height=0.35\textheight,angle=90]{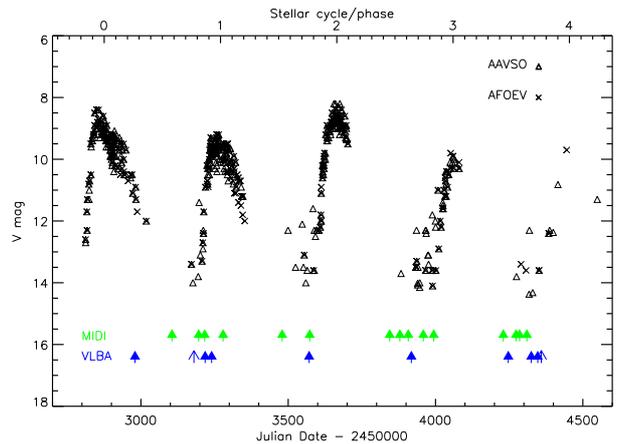}
 \caption{Visual light curve of RR~Aql based on data from the AAVSO and AFOEV 
databases as a function of Julian Date and stellar cycle/phase. 
The green arrows indicate the dates of our VLTI/MIDI observations. 
The blue arrows indicate the dates of our VLBA observations. 
Here, the full arrows denote observations of SiO maser emission 
and simple arrows observations of H$_2$O maser emission.
}
         \label{LightCurve}
\end{figure}
   \begin{table*}[tbp]
      \caption[]{VLTI/MIDI observation of RR~Aql.}
       \label{observations}
\centering 
     \begin{tabular}{lrrrrlrllrrrr}
            \hline
		\hline
            Ep.&date&Time&JD&$\Phi$$_{vis}$&Conf.&$B$& Disp.&BC&$B_p$ &P.A.&Seeing&$\tau$$_0$\\
&& [UTC]&&&&&Elem.&&[m]&[deg]&[$''$]&[msec]\\
&DD/MM/YYYY&&&&&&&&&&&\\\hline
A & 09/04/2004 & 07:26 & 2453105 & 0.58 & U2-U3 & 47m & Prism & HS & 33.57 & 3.21 & 0.45 & 7.4 \\
A & 10/04/2004 & 09:05 & 2453106 & 0.58 & U2-U3 & 47m & Prism & HS & 37.17 & 26.09 & 0.76 & 7.8 \\
A &	     & 09:52 & 2453106 & 0.58 & U2-U3 & 47m & Prism & HS & 39.86 & 33.48 & 0.64 & 8.2 \\
B & 09/07/2004 & 08:25 & 2453196 & 0.81 & U2-U3 & 47m & Prism & HS & 45.23 & 44.83 & 0.47 & 4.8 \\
C & 28/07/2004 & 08:09 & 2453215 & 0.86 & U2-U3 & 47m & Prism & HS & 42.54 & 41.57 & 1.36 & 1.0 \\
C & 29/07/2004 & 02:05 & 2453216 & 0.86 & U2-U3 & 47m & Prism & HS & 37.83 & 28.16 & 1.66 & 1.0 \\
C &	     & 03:33 & 2453216 & 0.86 & U2-U3 & 47m & Prism & HS & 42.84 & 39.49 & 0.75 & 2.3 \\
C &          & 04:55 & 2453216 & 0.86 & U2-U3 & 47m & Prism & HS & 45.99 & 44.67 & 1.04 & 1.6 \\
C &          & 06:33 & 2453216 & 0.86 & U2-U3 & 47m & Prism & HS & 46.16 & 45.71 & 3.38 & 0.5 \\
C &          & 07:04 & 2453216 & 0.86 & U2-U3 & 47m & Prism & HS & 45.32 & 44.92 & 2.68 & 0.7 \\
C & 01/08/2004 & 00:12 & 2453219 & 0.87 & U2-U3 & 47m & Prism & HS & 33.76 & 6.86 & 0.73 & 3.1 \\
D & 18/04/2005 & 10:30 & 2453479 & 1.53 & U2-U4 & 89m & Prism & HS & 88.37 & 81.72 & 0.76 & 3.5 \\
D & 19/04/2005 & 07:53 & 2453480 & 1.53 & U2-U4 & 89m & Prism & HS & 61.36 & 76.41 & 0.83 & 2.3 \\
E & 20/07/2005 & 03:51 & 2453572 & 1.76 & U1-U4 & 130m & Prism & HS & 119.63 & 59.78 & 0.71 & 1.9 \\
E &          & 06:16 & 2453572 & 1.76 & U1-U4 & 130m & Prism & HS & 128.64 & 63.28 & 0.81 & 1.6 \\
E & 22/07/2005 & 05:10 & 2453574 & 1.77 & U2-U3 & 47m & Prism & HS & 45.65 & 44.12 & 1.00 & 1.4 \\
E &          & 07:56 & 2453574 & 1.77 & U2-U3 & 47m & Prism & HS & 44.44 & 43.96 & 1.11 & 1.2 \\
F & 18/04/2006 & 09:38 & 2453844 & 2.45 & D0-G0 & 32m & Prism & HS & 28.99 & 70.01 & 1.08 & 2.1 \\
F & 19/04/2006 & 08:32 & 2453845 & 2.46 & D0-G0 & 32m & Prism & HS & 24.74 & 65.81 & 0.80 & 6.0 \\
F &          & 09:21 & 2453845 & 2.46 & D0-G0 & 32m & Prism & HS & 28.21 & 69.32 & 0.78 & 6.3 \\
G & 21/05/2006 & 05:53 & 2453877 & 2.54 & E0-G0 & 16m & Prism & HS & 10.94 & 62.10 & 0.56 & 4.0 \\
G & 25/05/2006 & 06:35 & 2453881 & 2.55 & A0-G0 & 64m & Prism & HS & 53.17 & 67.77 & 0.55 & 5.1 \\
G &          & 08:46 & 2453881 & 2.55 & A0-G0 & 64m & Prism & HS & 63.93 & 72.67 & 0.44 & 6.7 \\
H & 18/06/2006 & 03:58 & 2453905 & 2.61 & D0-G0 & 32m & Prism & HS & 21.37 & 61.34 & 1.29 & 1.5 \\
H &          & 04:41 & 2453905 & 2.61 & D0-G0 & 32m & Prism & HS & 25.09 & 66.21 & 1.27 & 1.6 \\
H & 18/06/2006 & 05:39 & 2453905 & 2.61 & D0-G0 & 32m & Prism & HS & 29.05 & 70.06 & 1.00 & 2.1 \\
H & 20/06/2006 & 05:49 & 2453907 & 2.61 & A0-G0 & 64m & Prism & HS & 59.89 & 70.81 & 1.05 & 3.9 \\
H &          & 06:33 & 2453907 & 2.61 & A0-G0 & 64m & Prism & HS & 63.00 & 72.15 & 0.88 & 4.5 \\
H & 21/06/2006 & 07:08 & 2453908 & 2.62 & E0-G0 & 16m & Prism & HS & 16.00 & 72.76 & 1.01 & 2.2 \\
H &          & 09:36 & 2453908 & 2.62 & E0-G0 & 16m & Prism & HS & 13.21 & 70.37 & 0.77 & 2.9 \\
H & 23/06/2006 & 03:20 & 2453910 & 2.62 & E0-G0 & 16m & Prism & HS & 9.84 & 58.42 & 0.76 & 3.9 \\
H &          & 04:07 & 2453910 & 2.62 & E0-G0 & 16m & Prism & HS & 11.97 & 64.86 & 0.67 & 4.4 \\
I & 08/08/2006 & 05:12 & 2453956 & 2.74 & A0-G0 & 64m & Prism & HS & 61.41 & 72.62 & 1.02 & 2.3 \\
I &          & 05:57 & 2453956 & 2.74 & A0-G0 & 64m & Prism & HS & 56.88 & 71.58 & 1.10 & 2.2 \\
I & 09/08/2006 & 06:39 & 2453957 & 2.74 & A0-G0 & 64m & Prism & HS & 50.33 & 69.49 & 1.47 & 1.5 \\
I & 10/08/2006 & 03:07 & 2453958 & 2.74 & D0-G0 & 32m & Prism & HS & 31.37 & 72.03 & 1.43 & 2.6 \\
I &          & 04:31 & 2453958 & 2.74 & D0-G0 & 32m & Prism & HS & 31.65 & 72.90 & 1.67 & 2.5 \\
I &          & 05:45 & 2453958 & 2.74 & D0-G0 & 32m & Prism & HS & 28.72 & 71.74 & 1.43 & 3.2 \\
I & 11/08/2006 & 02:44 & 2453959 & 2.74 & D0-G0 & 32m & Prism & HS & 30.76 & 71.51 & 1.43 & 2.3 \\
I & 13/08/2006 & 04:05 & 2453961 & 2.75 & E0-G0 & 16m & Prism & HS & 15.94 & 72.90 & 0.91 & 1.9 \\
I &          & 04:56 & 2453961 & 2.75 & E0-G0 & 16m & Prism & HS & 15.26 & 72.56 & 0.63 & 2.7 \\
I &          & 05:52 & 2453961 & 2.75 & E0-G0 & 16m & Prism & HS & 13.76 & 71.07 & 0.60 & 2.8 \\
J & 14/09/2006 & 01:01 & 2453993 & 2.83 & E0-G0 & 16m & Prism & HS & 15.82 & 72.29 & 0.85 & 2.3 \\
J & 16/09/2006 & 02:15 & 2453995 & 2.84 & A0-G0 & 64m & Prism & HS & 62.86 & 72.85 & 0.95 & 1.6 \\
J &          & 03:13 & 2453995 & 2.84 & A0-G0 & 64m & Prism & HS & 58.12 & 71.90 & 1.68 & 0.9 \\
K & 09/05/2007 & 08:53 & 2454230 & 3.43 & D0-H0 & 64m & Prism & HS & 61.43 & 71.46 & 1.12 & 1.3 \\
L & 22/06/2007 & 03:50 & 2454274 & 3.54 & G0-H0 & 32m & Prism & HS & 21.96 & 62.20 & 1.17 & 1.0 \\
L &          & 05:29 & 2454274 & 3.54 & G0-H0 & 32m & Prism & HS & 29.30 & 70.26 & 1.21 & 0.9 \\
M & 03/07/2007 & 07:25 & 2454285 & 3.57 & E0-G0 & 16m & Prism & HS & 15.51 & 72.73 & 1.07 & 2.8 \\
M &          & 07:37 & 2454285 & 3.57 & E0-G0 & 16m & Prism & HS & 15.31 & 72.59 & 0.75 & 3.9 \\
M & 04/07/2007 & 04:42 & 2454286 & 3.57 & G0-H0 & 32m & Prism & HS & 29.28 & 70.24 & 1.59 & 1.0 \\
M &          & 06:30 & 2454286 & 3.57 & G0-H0 & 32m & Prism & HS & 31.98 & 72.83 & 1.51 & 1.1 \\ \hline
\end{tabular}
\tablefoot{The table lists the epoch, 
the date, the time, the Julian Date (JD), the visual pulsation phase 
$\Phi$$_{vis}$, the baseline configuration, the ground length of the 
configuration, the dispersive element, the beam combiner BC, 
the projected baseline length $B_p$, the position angle on the 
sky P.A. (deg. east of north), the DIMM seeing (at 500 nm), and the 
coherence time $\tau$$_0$ (at 500 nm).}

\end{table*}

\section{Characteristics of RR~Aql}
 \label{sec:VLTI MIDI}
RR~Aql is an oxygen-rich Mira variable with spectral type M6e-M9 
\citep{Samus2004}. RR~Aql shows a strong silicate emission feature 
in its mid-infrared spectrum \citep{Lorenz-Martins2000}. 
In addition, it has relatively strong SiO, H$_2$O, and OH maser 
emission \citep{Benson1990}. \citet{Vlemmings2007} 
observed OH masers toward RR~Aql at five epochs with the VLBA. Based on 
these observations, the distance to RR~Aql is estimated to 
$D=633^{+214}_{-128}$\,pc (HIPPARCOS distance $D = 540$\,pc). 
Van Belle et al. (2002) obtained a $K$-band ($\lambda=2.2\,\mu$m, $\Delta\lambda=0.4\,\mu$m)
angular size with the Infrared Optical Telescope 
Array (IOTA). The uniform disk diameter (UD) of 
$\Theta_{\mathrm{UD}}=10.73\pm0.66$\,mas at phase $\Phi=0.48$, 
an effective temperature $T_{\mathrm{eff}}=2127\pm111$\,K, and a
bolometric flux 
$f_{\mathrm{bol}}=78.4\pm11.8\:10^{-8}$\,ergs cm$^{-2}$s$^{-1}$. 
\citet{Miyata2000} studied the dust around the star spectroscopically 
and obtained $F_{\mathrm{dust}}/F_{\mathrm{star}}$ at 10\,$\mu$m of 
1.49\,$\pm$\,0.02. 
\citet{Ragland2006} measured the closure phase with 
IOTA in the \textit{H}-band, and classified RR~Aql as a target with no 
detectable asymmetries. The IRAS flux at 12\,$\mu$m is 332\,Jy. The light curve 
in the \textit{V} band varies from $\sim$8~mag at maximum light to $\sim$14~mag 
at minimum light. \citet{Whitelock2000} shows that the \textit{K} 
magnitude of RR~Aql varies from $\sim$0.0~mag at maximum light to $\sim$1.0~mag at 
minimum light. RR~Aql is pulsating with a period of $P=394.78$ days 
\citep{Samus2004} and the Julian Date of the last maximum brightness 
is $T_0=245 2875.4$ (\citep{Pojmanski2005}. The period of pulsation 
visually corresponds to recent values from the 
AAVSO\footnote{http://www.aavso.org} and 
AFOEV\footnote{http://cdsweb.u-strasbg.fr/afoev} databases. 
Figure ~\ref{LightCurve} shows the visual light curve of RR~Aql based on 
values from the AAVSO and AFOEV databases as a function of Julian day 
and stellar phase. 

\section{VLTI/MIDI Observations and data reduction}
\begin{figure}
\includegraphics[height=0.35\textheight,angle=90]{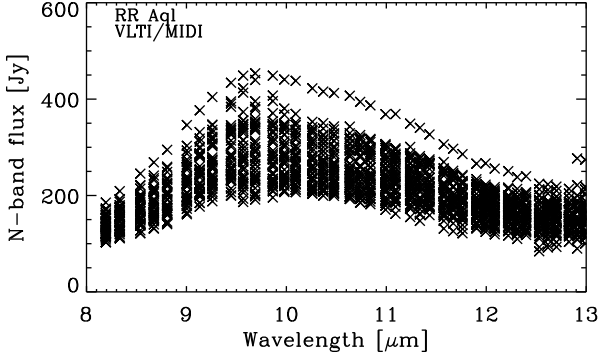}
 \caption{Calibrated RR~Aql MIDI flux spectra as a function of wavelength. 
For clarity, the error bars are omitted in the plot.}
\label{fig:phot}
\end{figure}
\begin{figure}
\centering
  \includegraphics[height=0.35\textheight,angle=90]{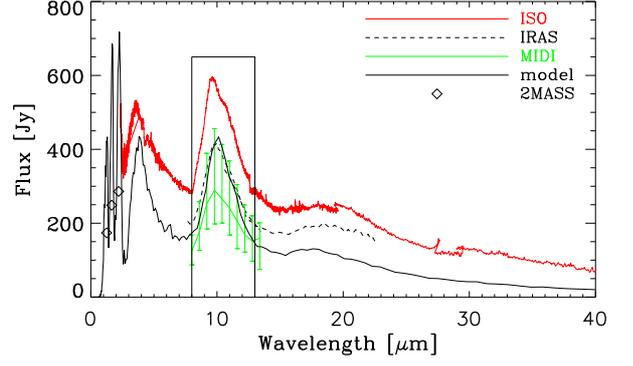}

  \includegraphics[height=0.35\textheight,angle=90]{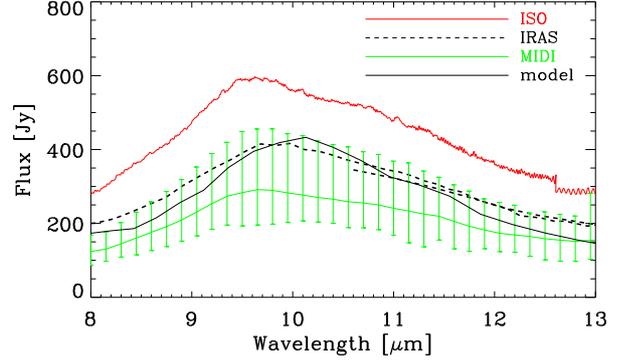}
 \caption{RR~Aql flux spectrum as a function of wavelength 
from 2.4\,$\mu$m to 40\,$\mu$m (top). The lines represent the flux spectra 
from ISO-SWS observations covering wavelengths from 2.5\,$\mu$m 
to $\sim$240\,$\mu$m  
(solid red thick line), IRAS-LRS observations covering
$\sim$ 7.7\,$\mu$m to $\sim$ 23\,$\mu$m 
(dashed thin black line), and the mean of our \textit{N}-band 
MIDI measurements
(solid thin green line). Here, the vertical bars span the maximum and 
minimum values measured. The diamonds denote 2MASS measurements 
at 1.25\,$\mu$m, 1.65\,$\mu$m, and 2.2\,$\mu$m. The solid black line indicates 
our atmosphere and dust shell model as explained in 
Sect.~\protect\ref{sec:results}. The bottom plot shows an enlarged 
segment of the plot in the MIDI wavelength range of 8--13\,$\mu$m.}
\label{iso_iras}
\end{figure}
We obtained 57 spectrally-dispersed mid-infrared interferometric 
observations of RR~Aql with the VLTI/MIDI instrument between 
Apr 9, 2004 and Jul 28, 2007. MIDI - the mid-infrared interferometric 
instrument \citep{Leinert2004} combines the beams from two telescopes 
of the VLTI \citep{Glindemann2003} and provides spectrally resolved 
visibilities in the 10\,$\mu$m window (\textit{N}-band, 8-13\,$\mu$m). 
To obtain dispersed photometric and interferometric signals, we used 
the PRISM as a dispersive element with a spectral resolution 
$R=\Delta\lambda/\lambda\,\sim\,30$. The beams were combined 
in \textit{High$\_$Sens} mode (HS). In this mode the photometric signal is 
observed after the interferometric signal.

The details of the observations and the instrumental settings are 
summarized in Table~\ref{observations}. All observations were executed 
in service mode using either the unit telescopes (UTs, 8.2m) or the 
auxiliary telescopes (ATs, 1.8m). 

We merged the MIDI data into 15 epochs, with a maximum time-lag between 
individual observations of five days for each epoch (1.3$\%$ of the 
pulsation period). For technical problems we had to eliminate two epochs. 
Fig.~\ref{LightCurve} shows the final 13 epochs in comparison to the light curve. The uncertainty in the allocation of 
the visual phase to our observations was estimated to $\sim$0.1. 
Our long-term VLBA 
monitoring of SiO and H$_2$O maser emission toward RR~Aql, 
will be presented in a forthcoming paper.

We used the MIA+EWS software package, version 1.6 (Jaffe, 
Koehler, et al\footnote{http://www.strw.leidenuniv.nl/$\sim$nevec/MIDI}) 
for the MIDI data reduction. This package includes two different methods, 
an incoherent method (MPIA software package MIA) that analyzes the 
powerspectrum of the observed fringe signal and a coherent integration 
method (EWS), which first compensates for optical path differences,
including both instrumental and atmospheric delays in each scan, and then 
coherently adds the fringes. We applied both methods to independently verify 
the data reduction results. The detector masks were calculated by the 
procedure of MIA, and were used for both the MIA and the EWS analysis. 
The obtained data reduction values correspond to each other, and we chose 
to use in the following the results derived from the EWS analysis, 
which offer error estimations. 

To account for instrumental visibility losses and to determine the absolute 
flux values, calibrator stars with known flux and diameter values were 
observed immediately before or after the science target. Our main calibrators were HD 169916 (period P73, P75), HD 146051 (P77), 
and HD 177716 (P79). Calibrated science target visibility spectra were calculated using the instrumental 
transfer function derived from all calibrator data sets taken during the 
same night with the same baseline and instrumental mode as our 
scientific target. The number of available transfer function measurements 
depends on the number of calibrator stars observed per specific night 
including those calibrators observed by other programs. The errors of the 
transfer functions are given by the standard deviation of all 
transfer function measurements per night. To estimate the uncertainty 
of the transfer function for nights when only one calibrator was available, 
we used typical values based on nights when many calibrator stars were 
observed. The final errors on the observed visibilities are mostly 
systematic, and include the error of the coherence factor of the 
science target and the calibrators, the adopted diameter errors, and 
the standard deviation of the transfer function over the night. 

The photometric spectrum was calibrated with one or two calibrators, 
which were observed close on the sky and in time compared to our science target. 
For most calibrator stars, absolutely calibrated spectra 
are available in \citet{Cohen1999}. For those calibrators where the 
absolutely calibrated spectrum was not directly available, we instead used 
a spectrum of a calibrator with a similar spectral type and similar effective 
temperature (see the instrument consortium's 
catalog\footnote{http://www.ster.kuleuven.ac.be/$\sim$tijl/MIDI$\_$calibration/mcc.txt}). 
The spectra of such calibrators were scaled with the IRAS flux 
at 12\,$\mu$m to the level of our calibrator. In addition, we verified that 
our synthetic spectra obtained by this procedure are valid by scaling
known spectra of two Cohen calibrators. In a few cases, when the 
atmospheric absorption was strongly affecting the spectra around 
9.5$\mu$m, we used another similar calibrator instead of the main 
calibrator observed in the same night with the same level of flux.
The ambient conditions for all the observations were carefully checked. 
If we were facing problems with clouds, constraints due to wind, 
significant differences in seeing, humidity, coherence time, and airmass 
between the science target and the corresponding calibrator, the photometry 
was omitted from the analysis. 
\begin{figure*}
\includegraphics[height=0.35\textheight,angle=90]{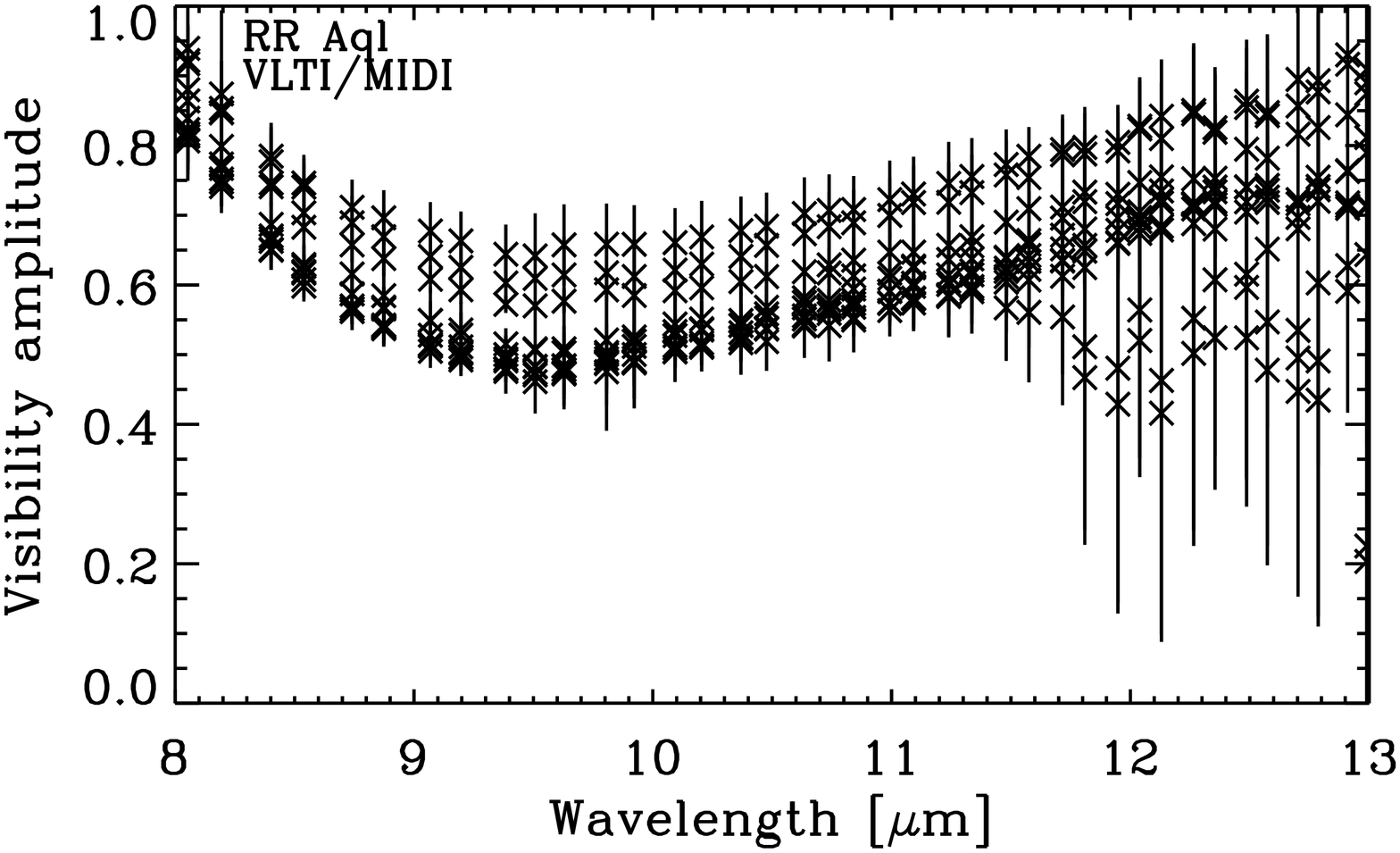}
\includegraphics[height=0.35\textheight,angle=90]{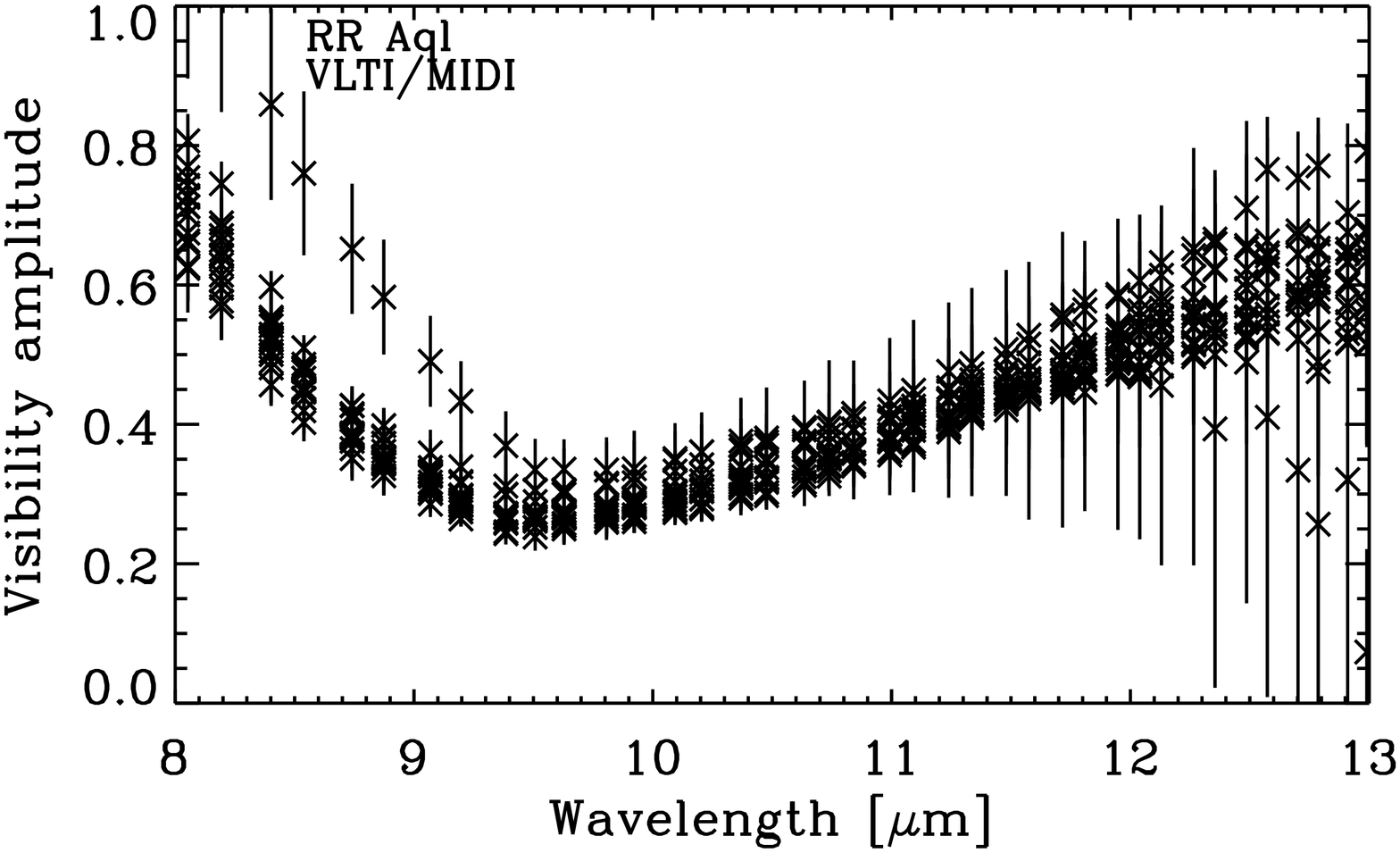}

\includegraphics[height=0.35\textheight,angle=90]{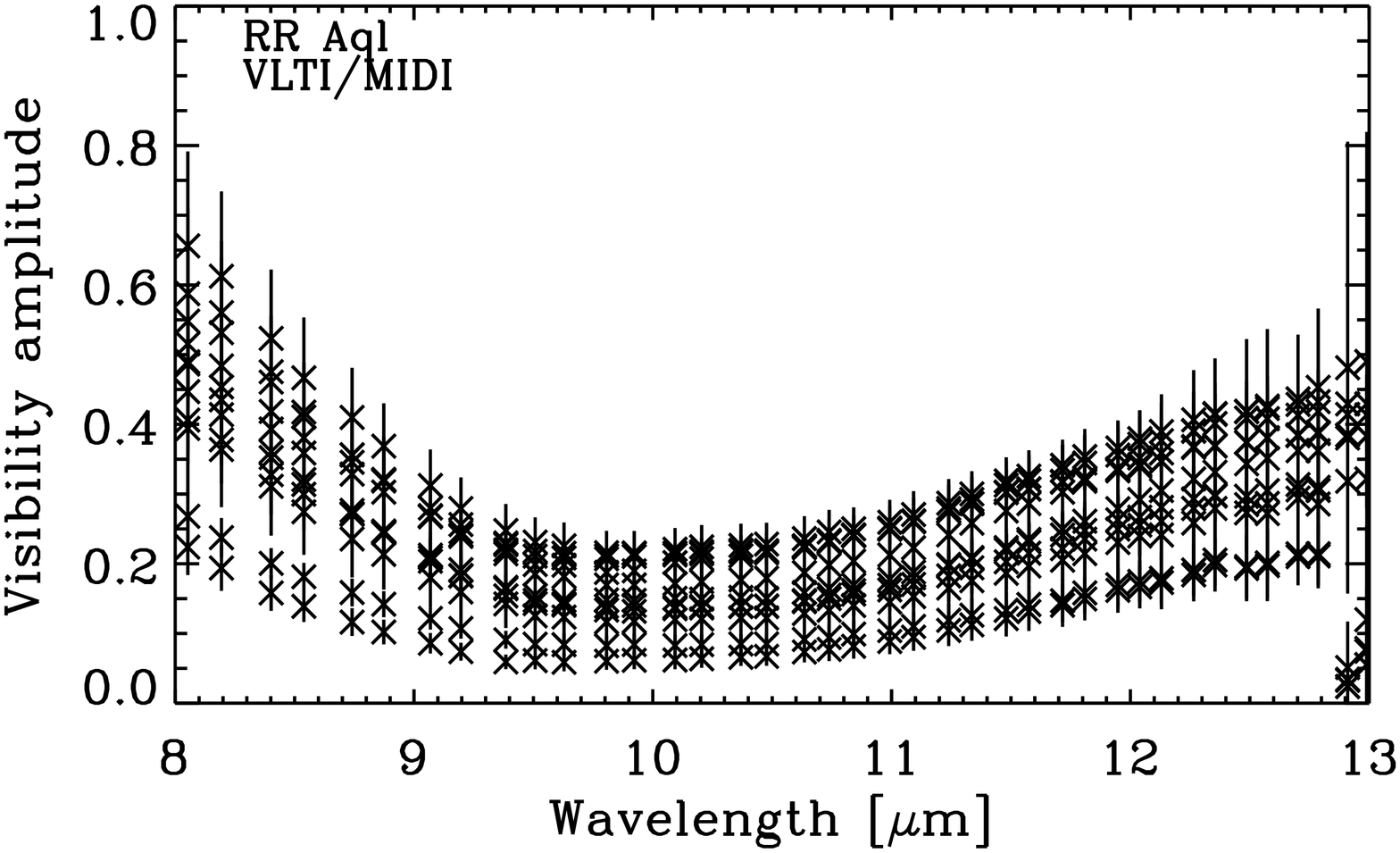}
\includegraphics[height=0.35\textheight,angle=90]{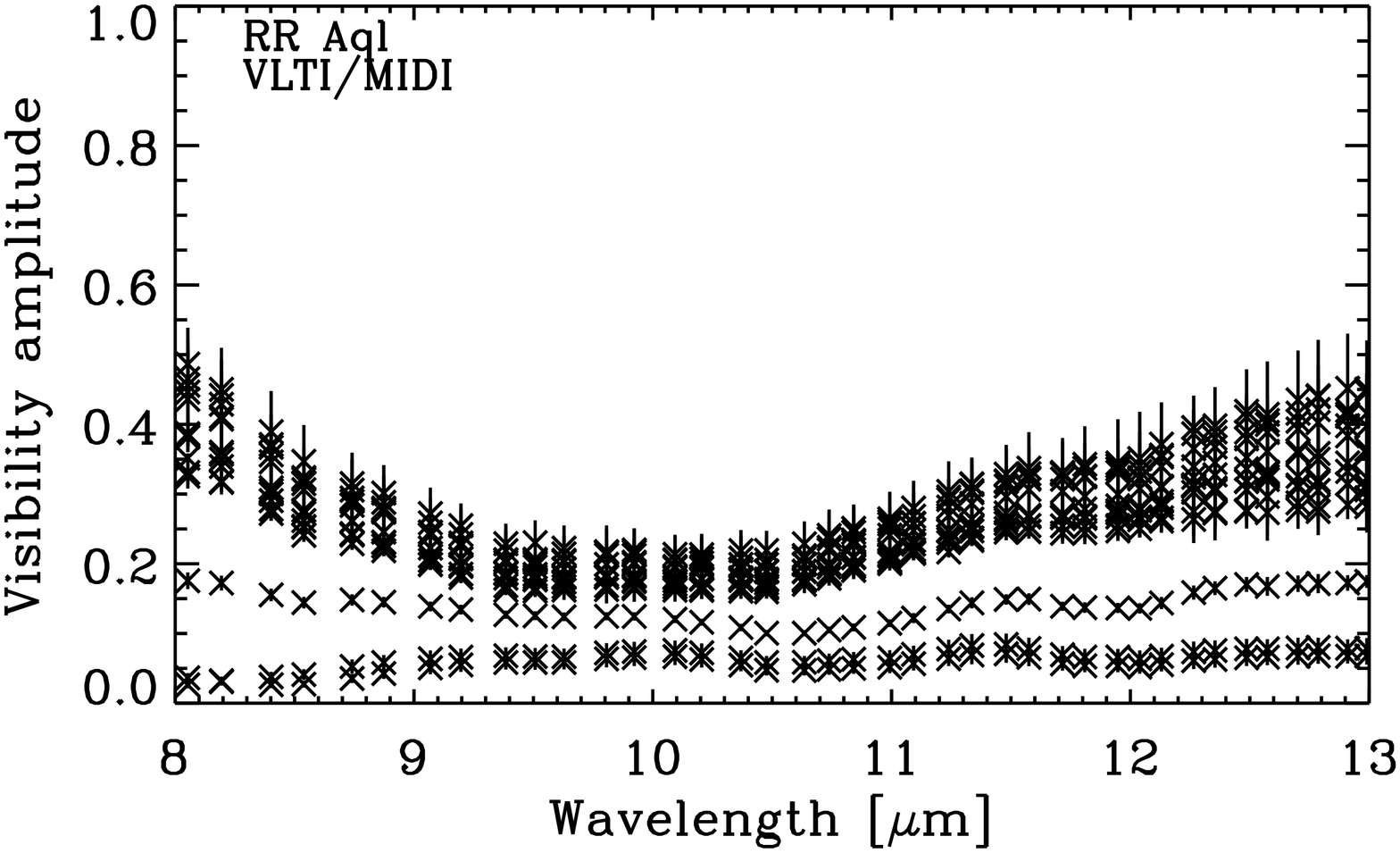}
 \caption{Calibrated RR~Aql MIDI visibility amplitudes as a function 
of wavelength. 
(top left) Observations executed at projected baselines ($B_p$) $<$ 20 m, 
(top right) 20 m $<$ $B_p$ $<$ 35 m, (bottom left) 35 m $<$ $B_p$ $<$ 50 m, 
and (bottom right) $B_p$ $>$ 50 m. }
         \label{fig:vis}
\end{figure*}
\section{MIDI results}
\label{sec:midiresults}
We observed RR~Aql at different phases over four pulsation cycles in order 
to monitor the mid-infrared photometry and visibility spectra. 
Here, we present and discuss the general properties of the data, followed
by an analysis of their variability as a function of phase and cycle
in Sects. \ref{sec:vismonitoring} and \ref{sec:fluxmonitoring}.

Figure~\ref{fig:phot}
shows all obtained calibrated 
photometry spectra as a function of wavelength. 
The MIDI flux measurements show a consistent shape exhibiting
an increase in the flux from $\sim$\,100--200\,Jy at 8\,$\mu$m
to a maximum near 9.8\,$\mu$m of $\sim$200--400\,Jy, and a decrease 
towards 13\,$\mu$m, where the flux values again reach values 
of $\sim$\,100--200\,Jy. The level of the flux spectrum
differs for individual measurements with a spread of $\sim$\,
100--200\,Jy. Figure~\ref{iso_iras} shows a comparison of the
mean of our MIDI flux measurements to measurements obtained
with the ISO and IRAS instruments. The shape of the flux curve
is consistent among the MIDI, ISO, and IRAS measurements. 
The level of the IRAS flux is within the range of our MIDI measurements. 
The level of the ISO flux is higher, which can most likely be explained
by the post-maximum phase of 0.16 of the ISO observations (1997-05-03) compared to our minimum to pre-maximum phases. The difference between the MIDI and ISO
flux level may also indicate a flux variation over different cycles.
Figure~\ref{iso_iras} also includes a model description of 
our MIDI data, which is explained below in Sect.~\ref{sec:results}.

Figure~\ref{fig:vis}
shows all obtained calibrated
visibility spectra, which are combined into four groups of different
projected baseline lengths ($B_p$) of $B_p$ $<$ 20 m, 
20 m $<$ $B_p$ $<$ 35 m, 35 m $<$ $B_p$ $<$ 50 m, and $B_p$ $>$ 50 m.
The visibility curves show a significant wavelength dependence with a 
steep decrease from 8\,$\mu$m to $\sim$ 9.5\,$\mu$m and a slow increase
in the 9.5\,$\mu$m to 13\,$\mu$m range.
The shape and absolute scale of the visibility function depends on the length 
of the projected baseline where a longer baseline results in a 
lower and flatter curve. 

As a first basic interpretation of the interferometric data,
we computed the corresponding uniform disk (UD) diameter and the
Gaussian FWHM for each
data set and spectral channel. 
This leads to a rough estimate of the characteristic size of the target 
at each wavelength. However, it should be mentioned that the true 
intensity distribution across the stellar disk is expected to be more 
complex than can be described by these elementary models. 
Figure~\ref{A} shows for the example of epoch A, and epoch I the flux, 
the visibility amplitude, the corresponding UD diameter, and
the corresponding Gaussian FWHM diameter as a function of wavelength.
The shape of the 
UD diameter and Gaussian FWHM functions show a steep increase 
from 8\,$\mu$m to $\sim$\,9.5\,$\mu$m by a factor of 2. A plateau appears 
between $\sim$\,9.5\,$\mu$m and 11.5\,$\mu$m, and the values are nearly constant from $\sim$\,11.5\,$\mu$m to 13\,$\mu$m.

\onlfig{5}{
\begin{figure*}
\centering
\includegraphics[height=0.3\textheight,angle=90]{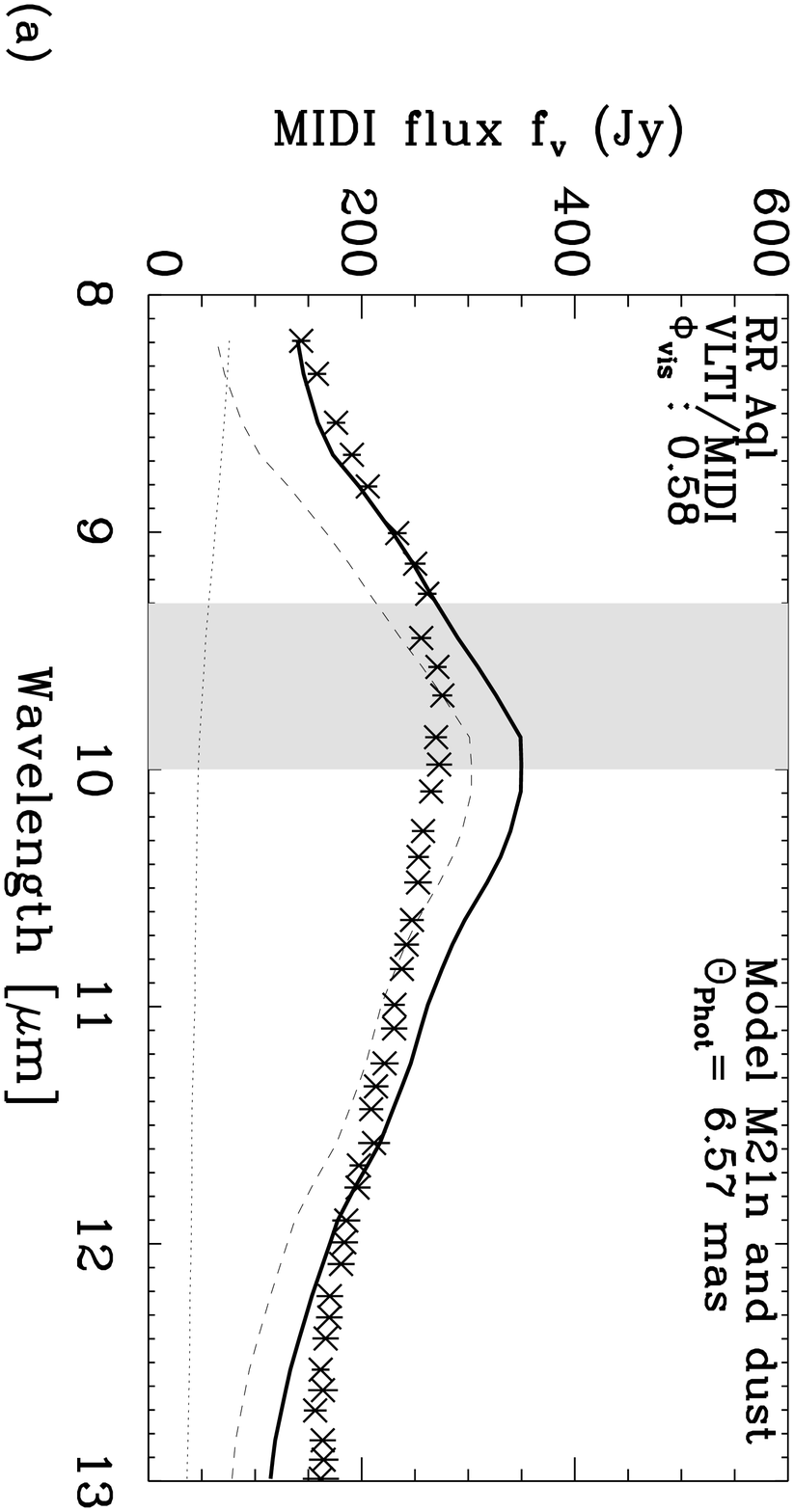}
\includegraphics[height=0.3\textheight,angle=90]{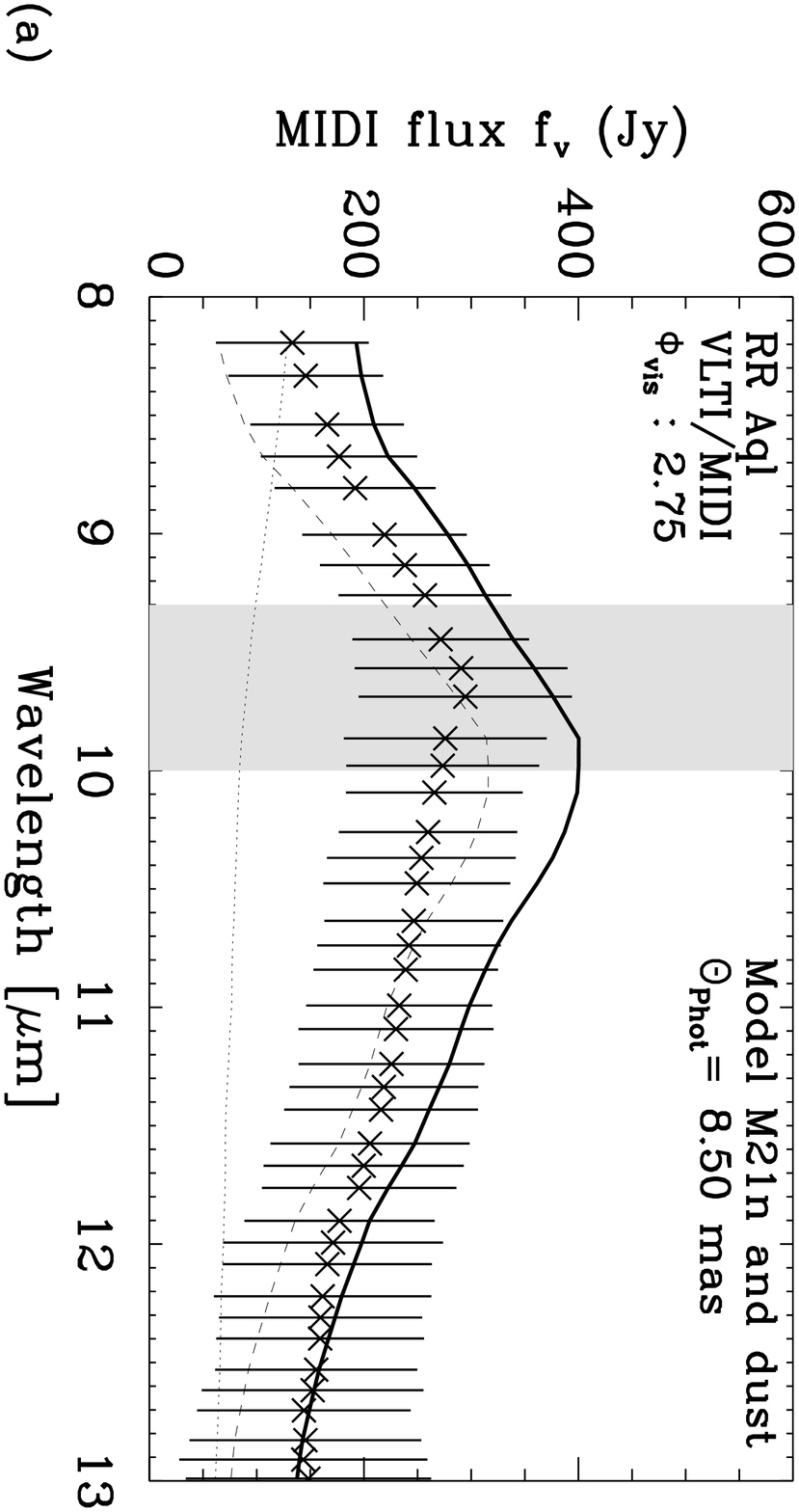}

\includegraphics[height=0.3\textheight,angle=90]{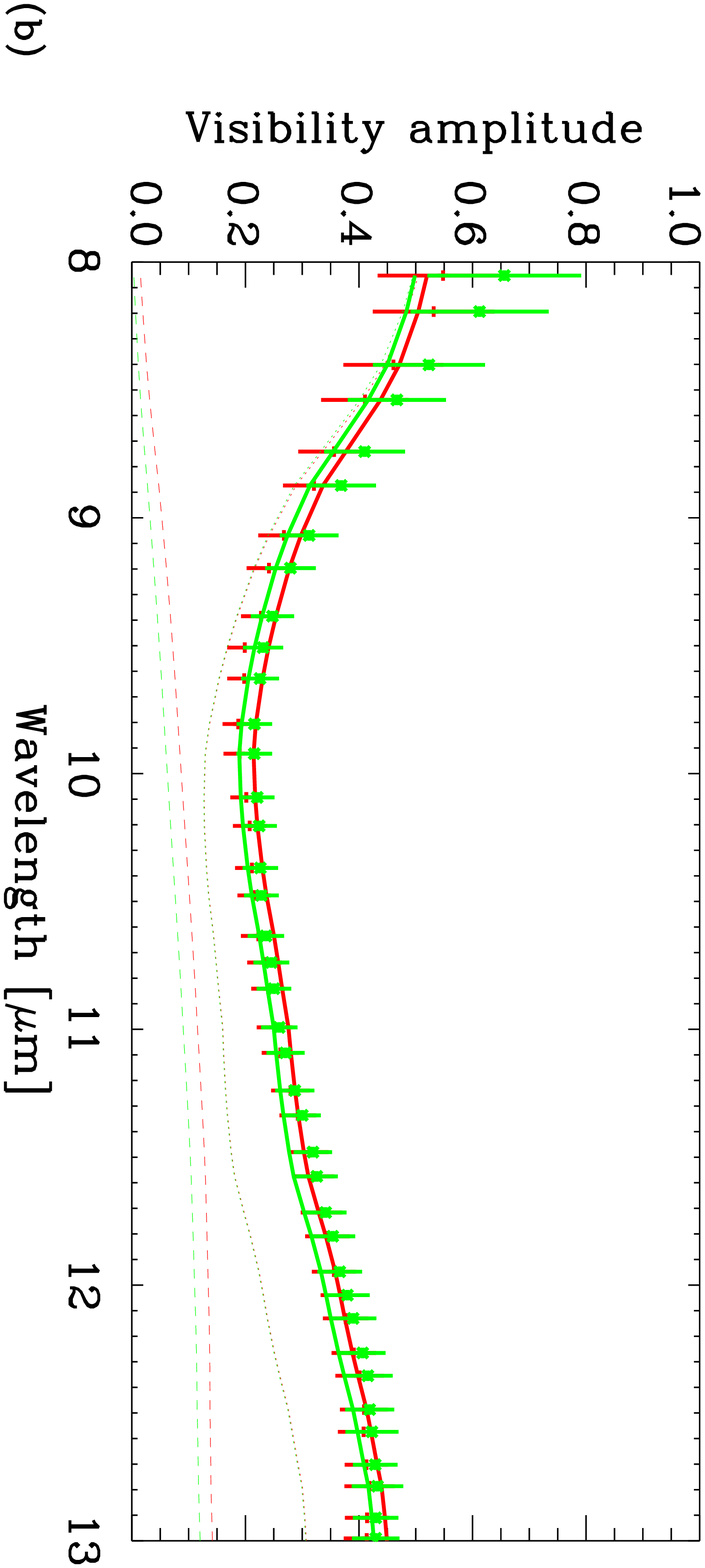}
\includegraphics[height=0.3\textheight,angle=90]{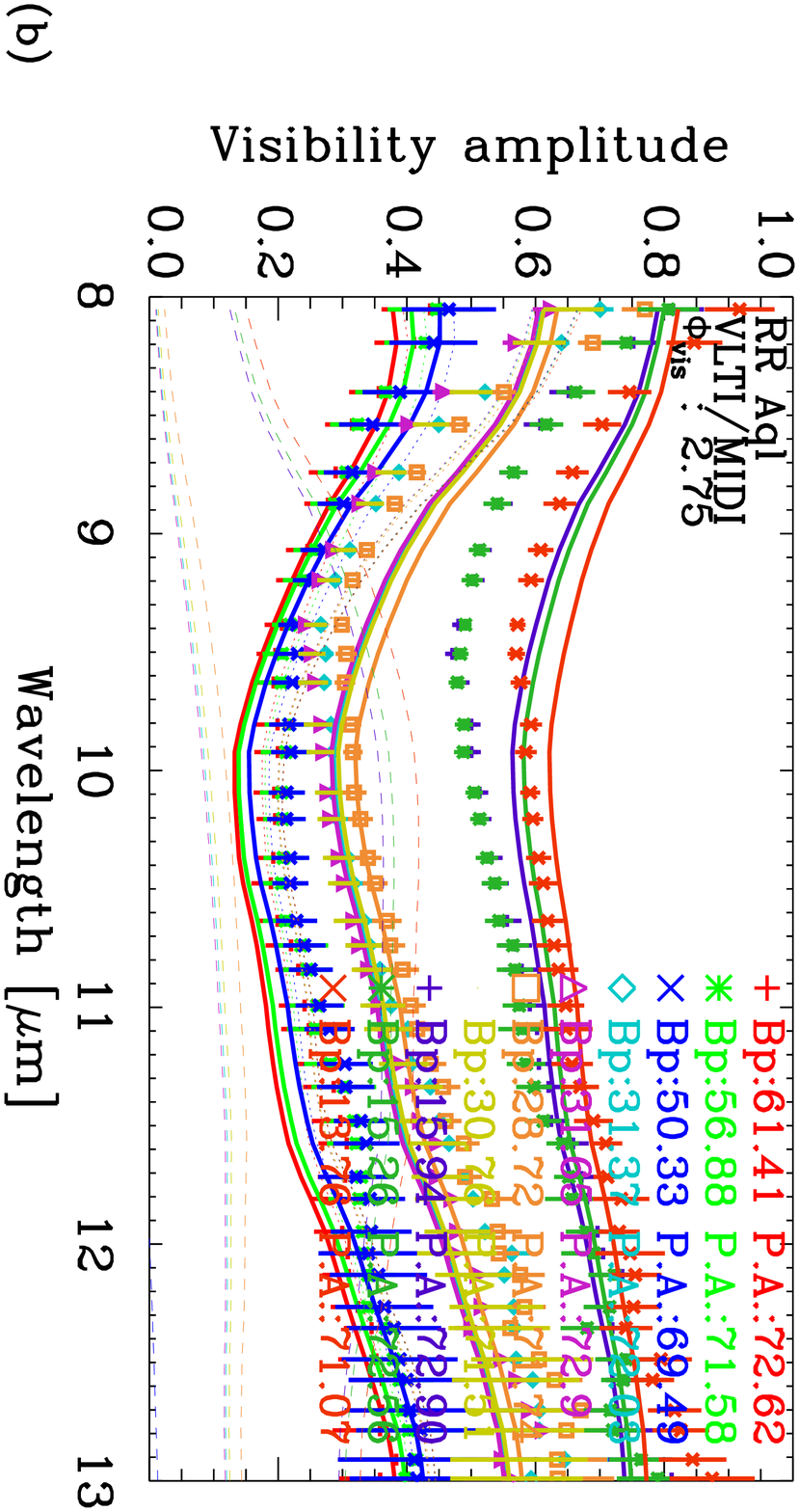}

\includegraphics[height=0.3\textheight,angle=90]{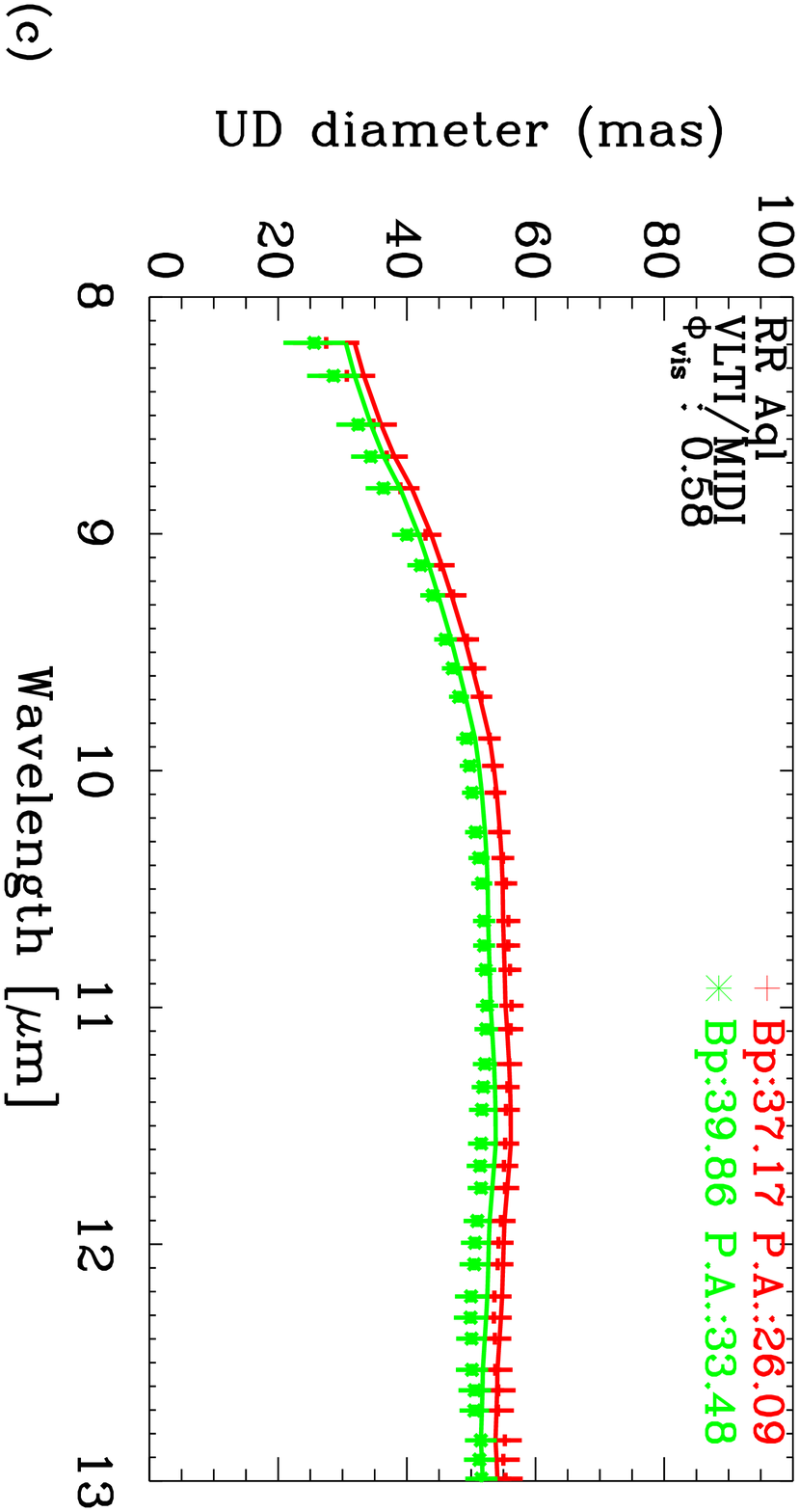}
\includegraphics[height=0.3\textheight,angle=90]{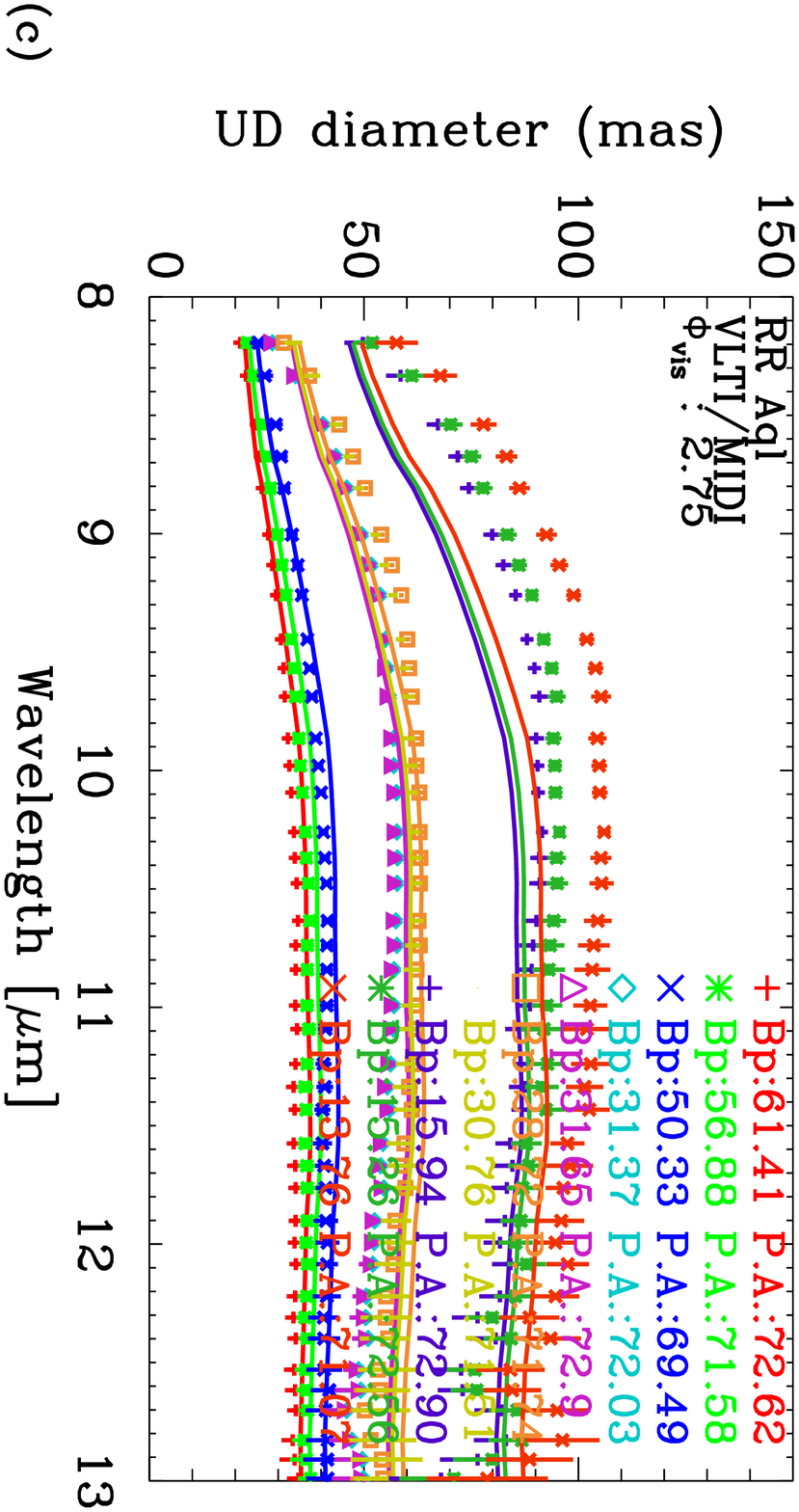}

\includegraphics[height=0.3\textheight,angle=90]{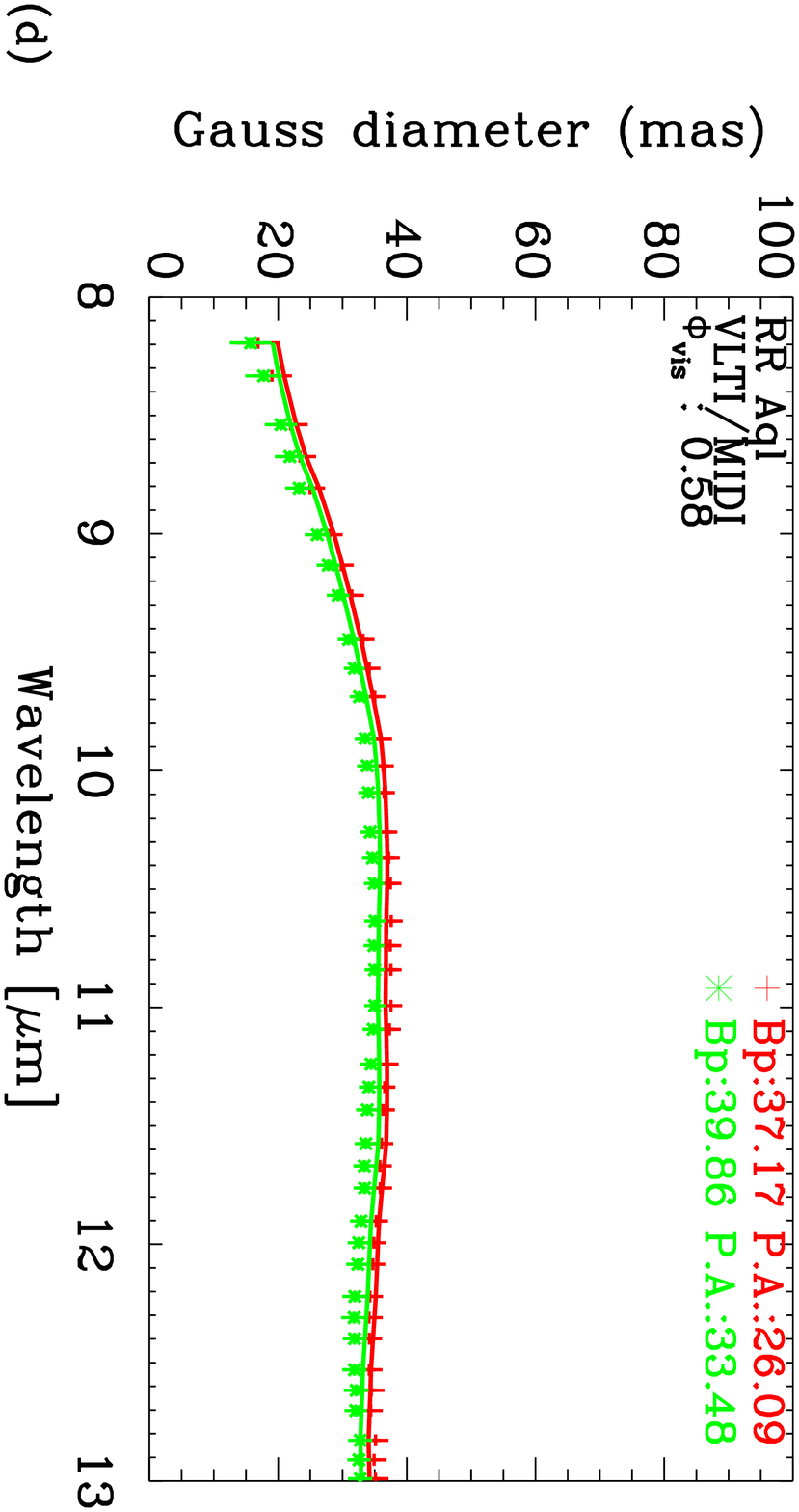}
\includegraphics[height=0.3\textheight,angle=90]{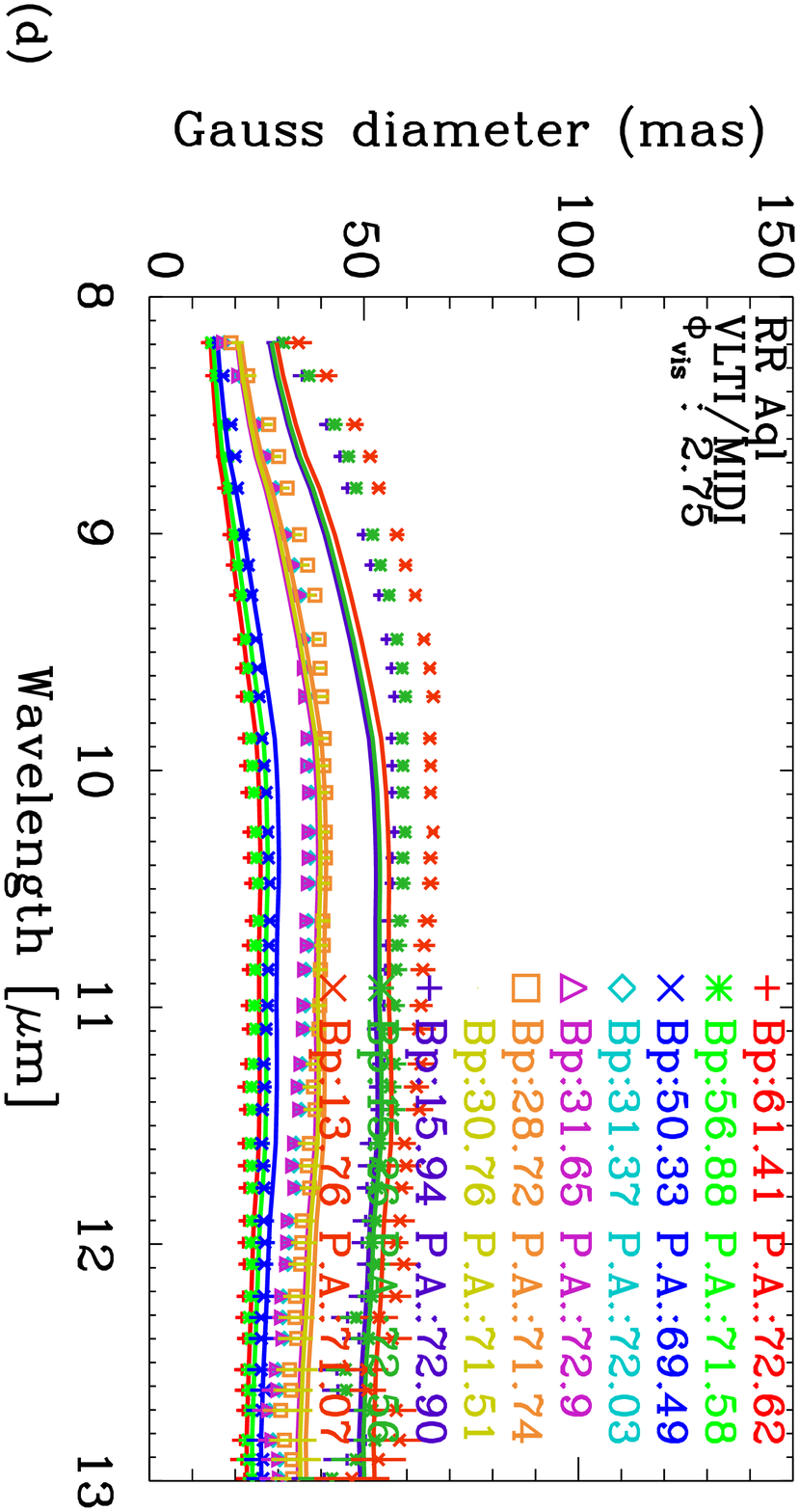}

\includegraphics[height=0.3\textheight,angle=90]{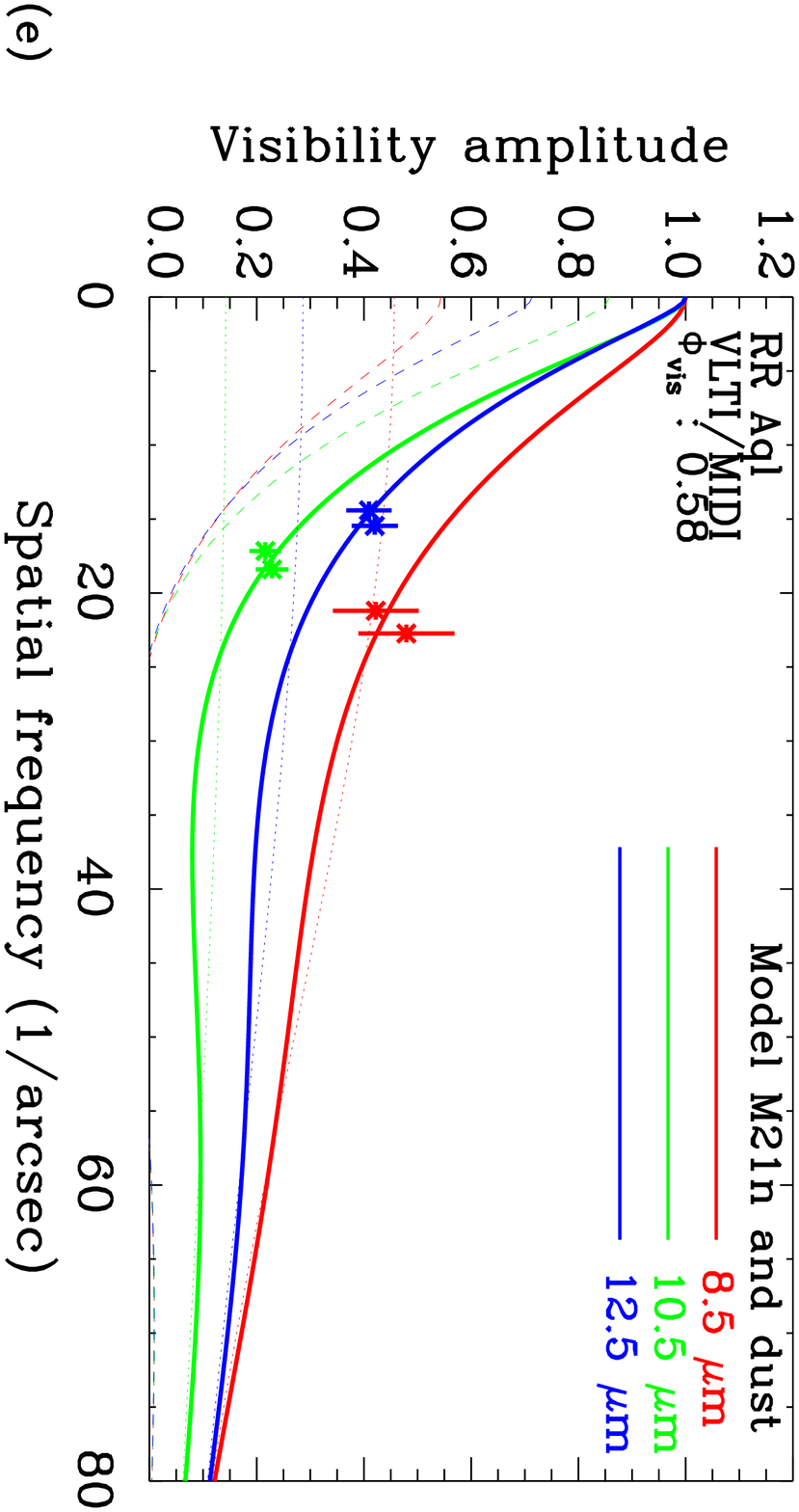}
\includegraphics[height=0.3\textheight,angle=90]{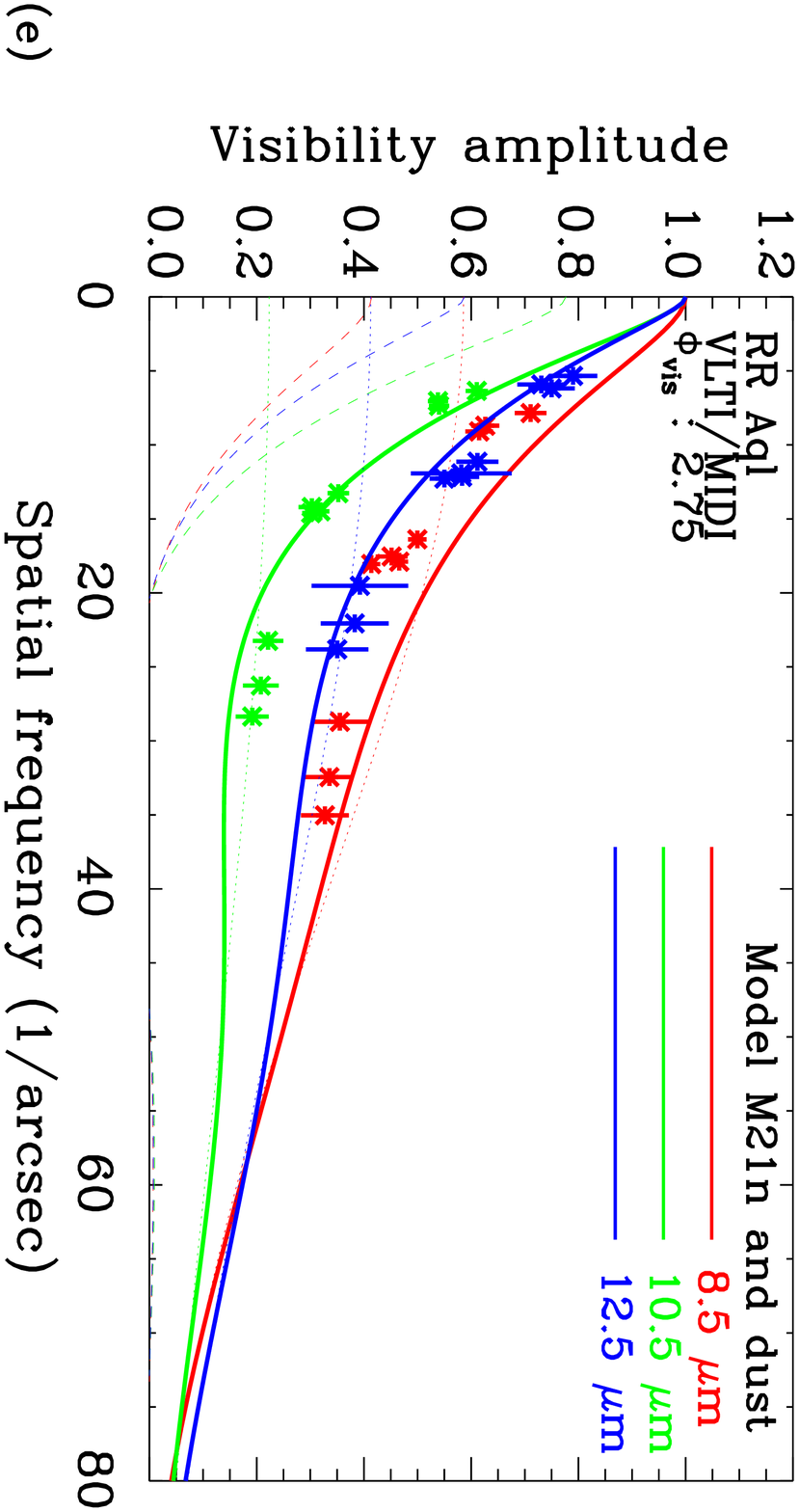}

\caption{VLTI/MIDI interferometry at 8-13$\mu$m of RR~Aql for the example
of epoch A (stellar phase 0.58) and epoch I (stellar phase 2.75) . The panels show 
(a) the flux, (b) the visibility amplitude, (c) the 
corresponding UD diameter, (d) the corresponding
Gaussian FWHM diameter as a function of wavelength.
The gray shading indicates the wavelength region around 9.5\,$\mu$m
that is affected by atmospheric absorption.
Panel (e) shows the visibility amplitude as a function 
of spatial frequency for three averaged bandpasses of 
8-9\,$\mu$m, 10-11\,$\mu$m, and 12-13\,$\mu$m.
The crosses with error bars denote the measured values.
The solid lines indicate our best-fitting model, as described
in Sect. \protect\ref{sec:modeling}. It consists of
a combination of a dust-free 
dynamic model atmosphere representing the central star
and a radiative transfer model representing the 
surrounding dust shell. The contributions of the 
stellar and dust components alone are indicated by the
dotted and the dashed lines, respectively.}
\label{A}
\end{figure*}
}

The shape of the 8--13\,$\mu$m flux spectrum with a maximum 
near 9.8\,$\mu$m is known to be a 
characteristic silicate emission feature \citep[e.g.][]{Little-Marenin1990,Lorenz-Martins2000}.
The corresponding drop in the visibility function 
between 8\,$\mu$m and 9.5\,$\mu$m
and increase in the UD diameter and Gaussian FWHM 
has also been shown to be a typical signature of a circumstellar dust 
shell that is
dominated by silicate dust \citep[cf., e.g.][]{Driebe2008,Ohnaka2008}. The opposite trend toward a broad spectrum
and visibility 
function in the wavelength range 8--13\,$\mu$m was interpreted as a 
dust shell 
consisting of Al$_{2}$O$_{3}$ dust \citep{Wittkowski2007} or silicate 
and Al$_2$O$_3$ dust \citep{Ohnaka2005}.

\begin{figure}
\includegraphics[height=0.35\textheight,angle=90]{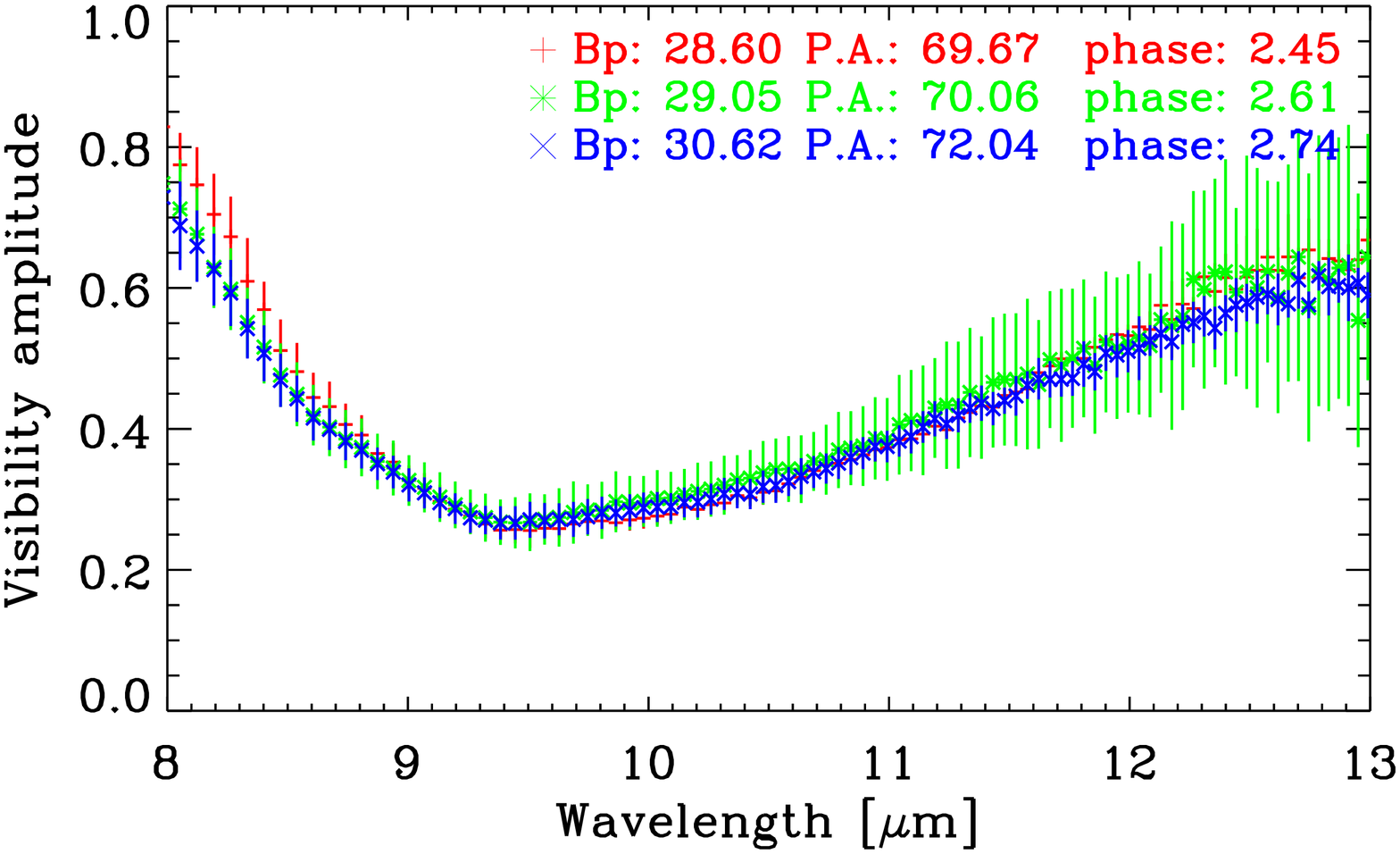}
\includegraphics[height=0.35\textheight,angle=90]{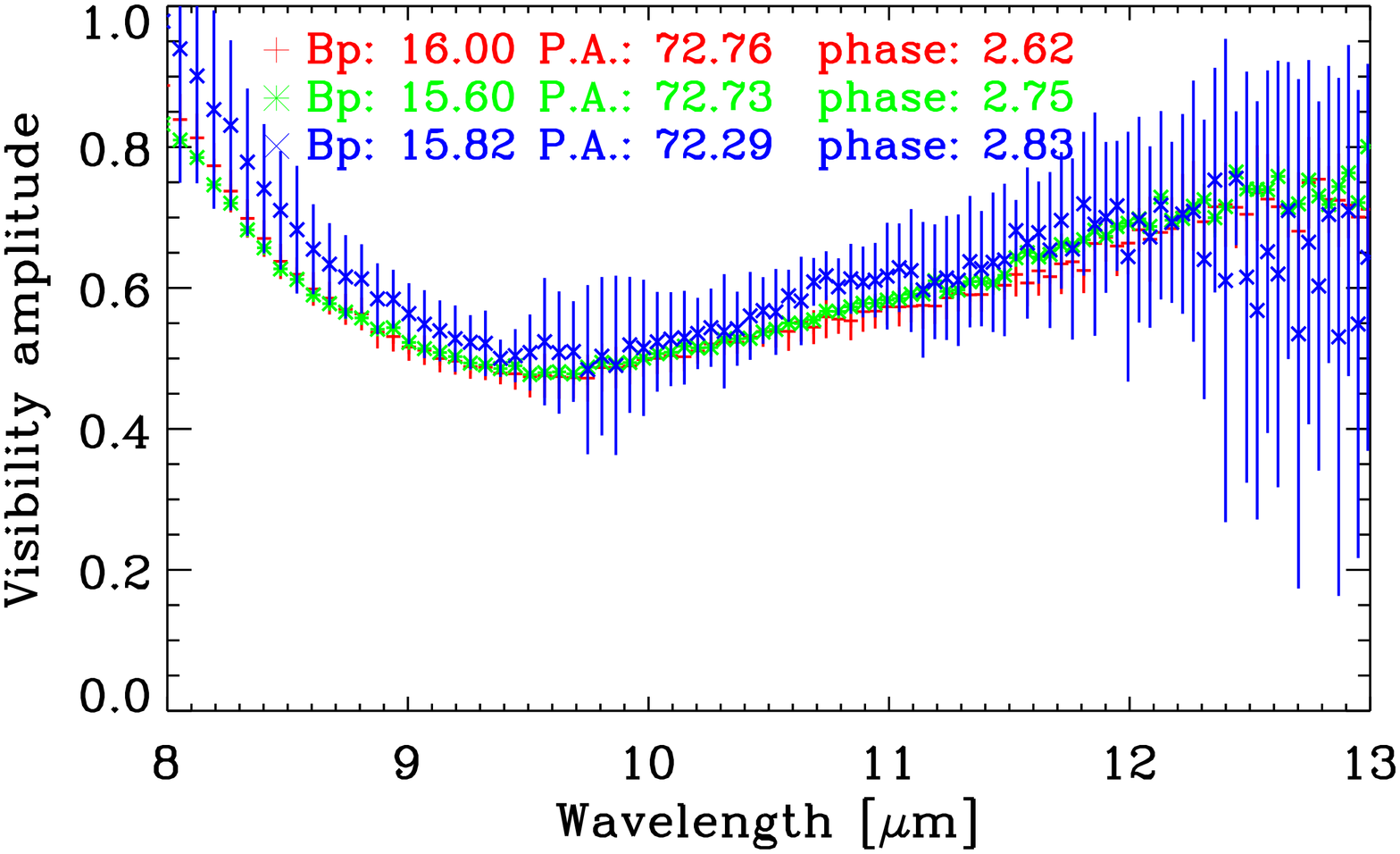}
 \caption{Calibrated MIDI visibility amplitudes
for different pulsation phases within the 
same cycle to investigate intracycle visibility variations. 
Each line represents a different pulsation phase within the same cycle
and is computed as an average of data obtained at the respective
phase ($\pm$ 0.15) and observed at similar projected baseline length
($B_p$ $\pm$ 10\%) and position angle (P.A. $\pm$ 10\%).
The top panel shows the example of pulsation phases 0.45, 0.61, and 0.74
of cycle 2 observed with a projected baseline length of $\sim$ 29\,m
and a position angle of $\sim$70$\deg$. The bottom panel shows the example of
pulsation phases 0.62, 0.75, and 0.83 of cycle 2 observed with a 
projected baseline length of $\sim$ 16\,m and a position angle 
of $\sim$72$\deg$.
The error bars are computed as the standard deviation of the
averaged visibilities.}
\label{fig:vis_ic_3}
\end{figure}

\begin{figure}
\includegraphics[height=0.35\textheight,angle=90]{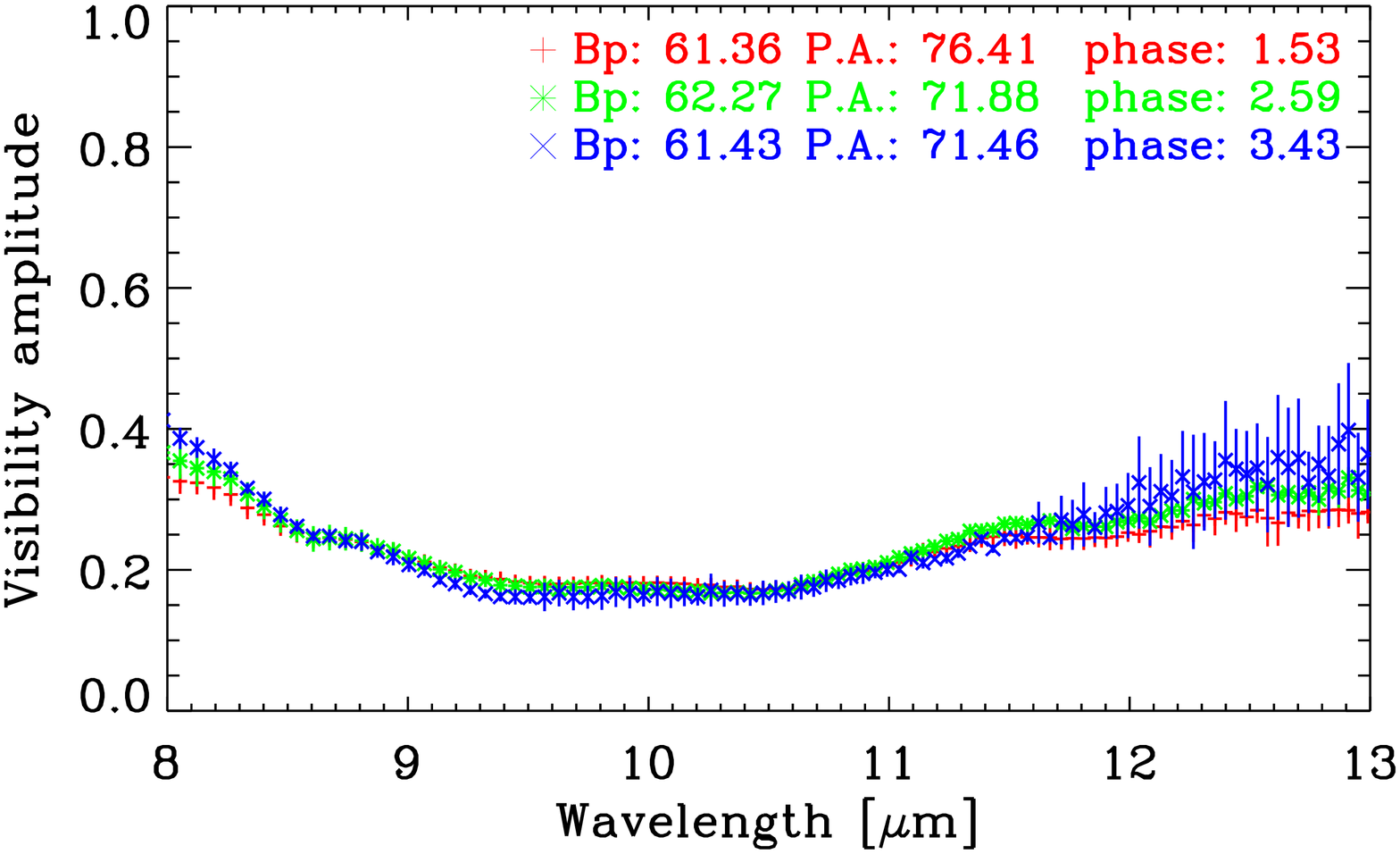}
\includegraphics[height=0.35\textheight,angle=90]{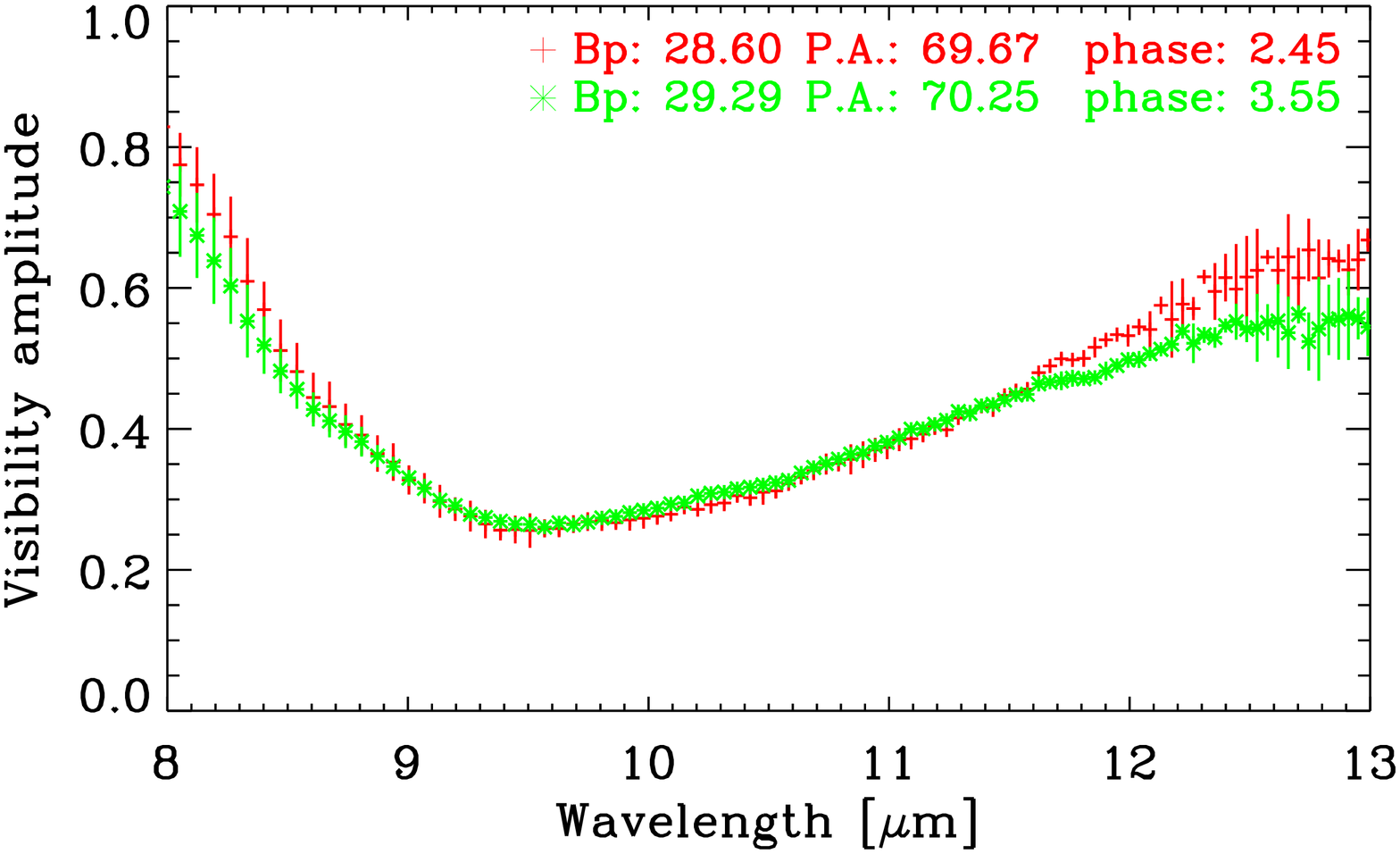}
 \caption{
As for Fig.~\protect\ref{fig:vis_ic_3}, but for the same pulsation phase
in consecutive pulsation cycles to investigate cycle-to-cycle
visibility variations. 
The top panel shows the example of pulsation phase $\sim$ 0.5
in three consecutive cycles 
observed with a projected baseline length of $\sim$ 61\,m
and a position angle $\sim$75$\deg$. The bottom panel shows the example of
pulsation phase 0.5 in 2 consecutive cycles observed with a
projected baseline length of $\sim$ 29\,m and a position angle $\sim$70$\deg$.}
\label{fig:vis_ctc_3}
\end{figure}

\subsection{Visibility monitoring}
\label{sec:vismonitoring}
We obtained a rich sample of MIDI data on RR~Aql, which covers a 
total of 4 pulsation cycles and pulsation phases between 0.45 and 0.85, 
i.e. minimum to pre-maximum phases (see Table~\ref{observations} and 
Fig.~\ref{LightCurve}).
For many different pulsation cycles and phases, we obtained
data at similar projected baseline lengths and position angles.
This gives us a unique opportunity to meaningfully compare interferometric 
data obtained at different pulsation phases and cycles. Since the
visibility depends on the probed point on the $uv$ plane, and thus
on the projected baseline length and position angle, visibility data
at different pulsation phases can be directly compared only if they
were obtained at the same, or very similar, point on the $uv$ plane.
For this reason, we combined individual observations
into groups of similar pulsation phases ($\Phi$$_\mathrm{vis}$ $\pm$ 0.15),
projected baseline lengths ($B_p$ $\pm$ 10 $\%$), and position 
angles (P.A. $\pm$ 10 $\%$).
The data within each group were averaged. The uncertainty of the averaged
visibility curves was estimated as the standard deviation
of the averaged visibilities. 

\subsubsection{Intracycle visibility monitoring}
\label{sec:intracycle}
To investigate intracycle visibility variations we compared data 
observed at different pulsation phases within the same cycle. 
Figure~\ref{fig:vis_ic_3} 
shows two examples of calibrated visibility curves,
where each line in the plot 
represents an average of a group of visibility data, as described above, 
(i.e. with similar $B_p$, P.A., and $\Phi$$_\mathrm{vis}$). The top panel
shows an example of 
observations within the second pulsation cycle at 
phases 2.45, 2.61, and 2.74 obtained with $B_p$ $\sim$ 30 m and 
P.A. $\sim$ 70$^\circ$. The bottom panel shows an example of 
observations at 
phases 2.62, 2.75, and 2.83 obtained with $B_p$ $\sim$ 16 m 
and P.A. $\sim$ 72$^\circ$. 
The data do not show any evidence of intracycle visibility variations
within the probed range of pulsation phases ($\sim$ 0.45 -- 0.85) and
within our visibility accuracies of about 
5--20\%.

\subsubsection{Cycle-to-cycle visibility monitoring}
We compared data observed at similar pulsation phases of different
consecutive cycles to investigate cycle-to-cycle visibility 
variations. 
Figure~\ref{fig:vis_ctc_3} shows two of these examples computed in
the same way as shown in Fig.~\ref{fig:vis_ic_3} of 
Sect.~\ref{sec:intracycle}, but (top panel) for minimum phases of 1.53, 2.59, 
and 3.43 in three consecutive cycles obtained with a projected 
baseline length of $\sim$ 62 m and P.A. $\sim$ 73$^\circ$, and (bottom panel)
for minimum phases 2.43, and 3.55 of two consecutive cycles
obtained with a projected baseline length of $\sim$ 29 m, 
and P.A. $\sim$ 70$^\circ$.
Here, the phases of the different cycles differ 
by up to $\sim$ 20\%. However, we showed in Sect.~\ref{sec:intracycle}
that there is no evidence of intracycle visibility variations,
so that the chosen phases can be compared well for different cycles.
As a result, Fig.~\ref{fig:vis_ctc_3} shows that our data do not exhibit 
any significant cycle-to-cycle visibility variation for minimum phases
over two or three cycles within our visibility accuracies of $\sim$ 5--20\%.

\begin{figure}
\includegraphics[height=0.35\textheight,angle=90]{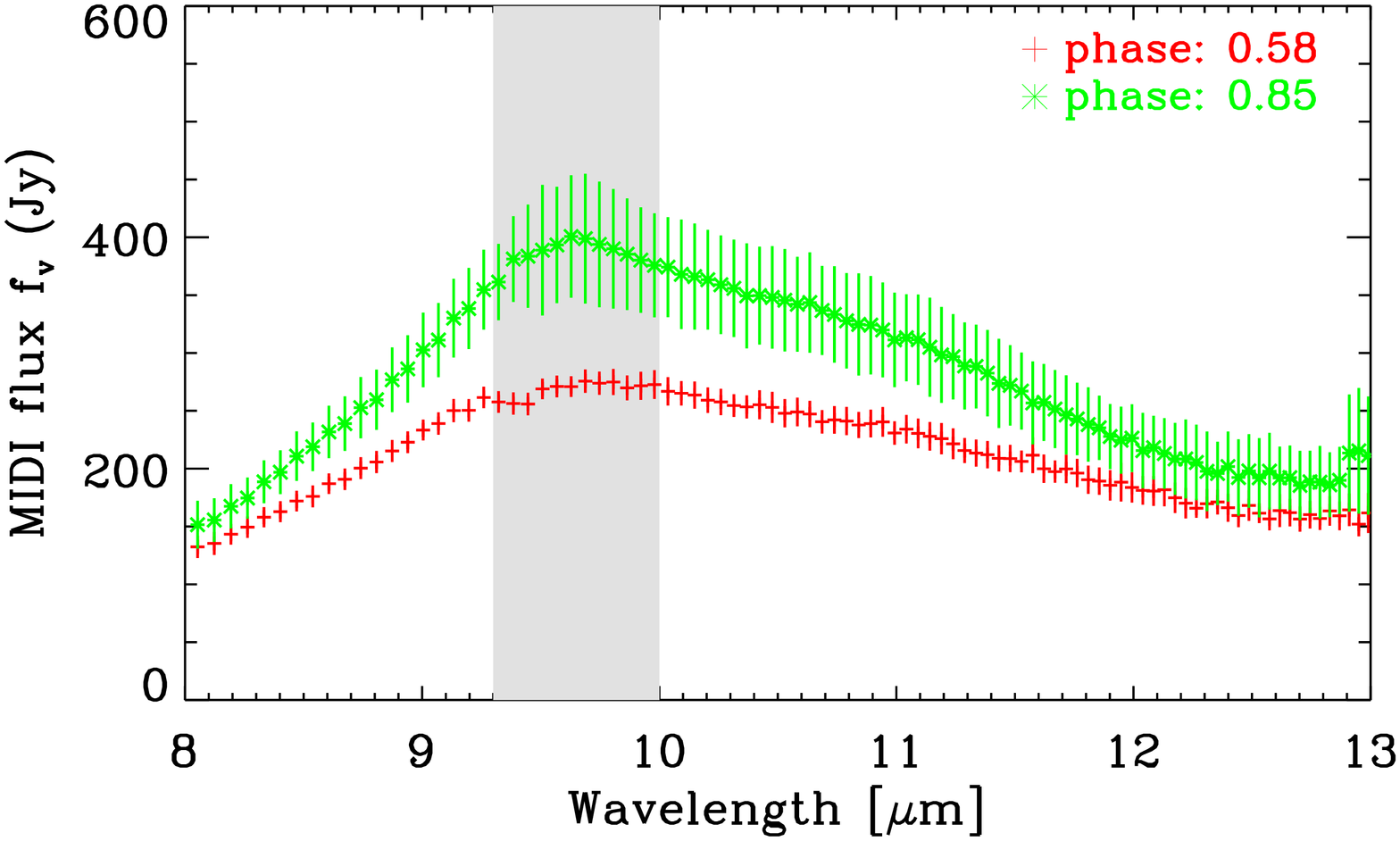}
\includegraphics[height=0.35\textheight,angle=90]{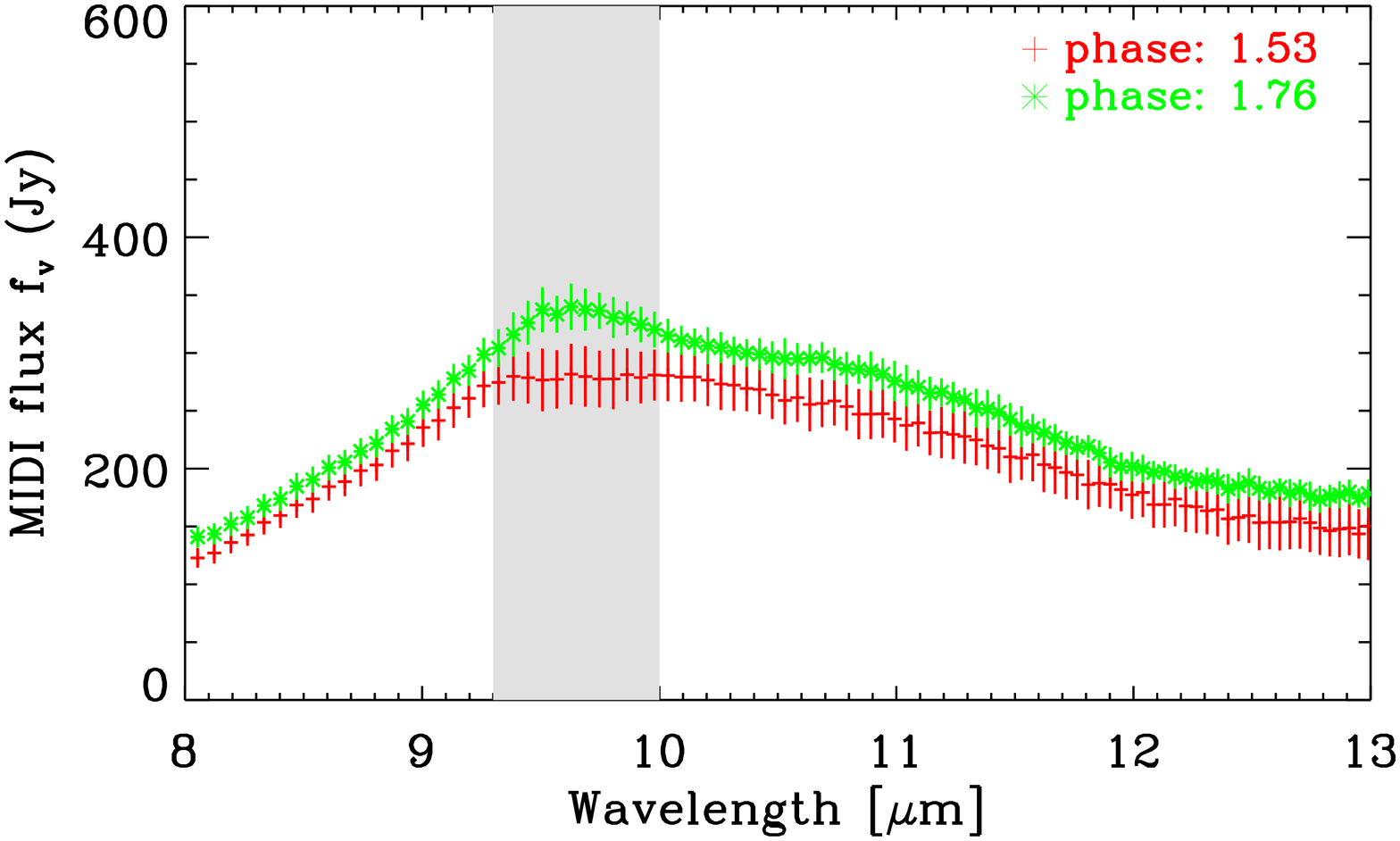}
 \caption{Calibrated MIDI flux spectrum 
for different pulsation phases within the
same cycle to investigate intracycle photometry variations. 
Each line represents a different pulsation phase within the same cycle
and is computed as an average of data obtained at the respective
phase ($\pm$ 0.15). The top panel shows the example of pulsation phases 
0.58 and 0.85 of cycle 0. The bottom panel shows the examples
of phases 0.53 and 0.76 of cycle 1. The error bars are computed as the 
standard deviation of the averaged photometry curves.
The gray shades denote zones
that are affected by atmospheric absorption.} 
\label{phot_ic_1}
\end{figure}

\begin{figure}
\includegraphics[height=0.35\textheight,angle=90]{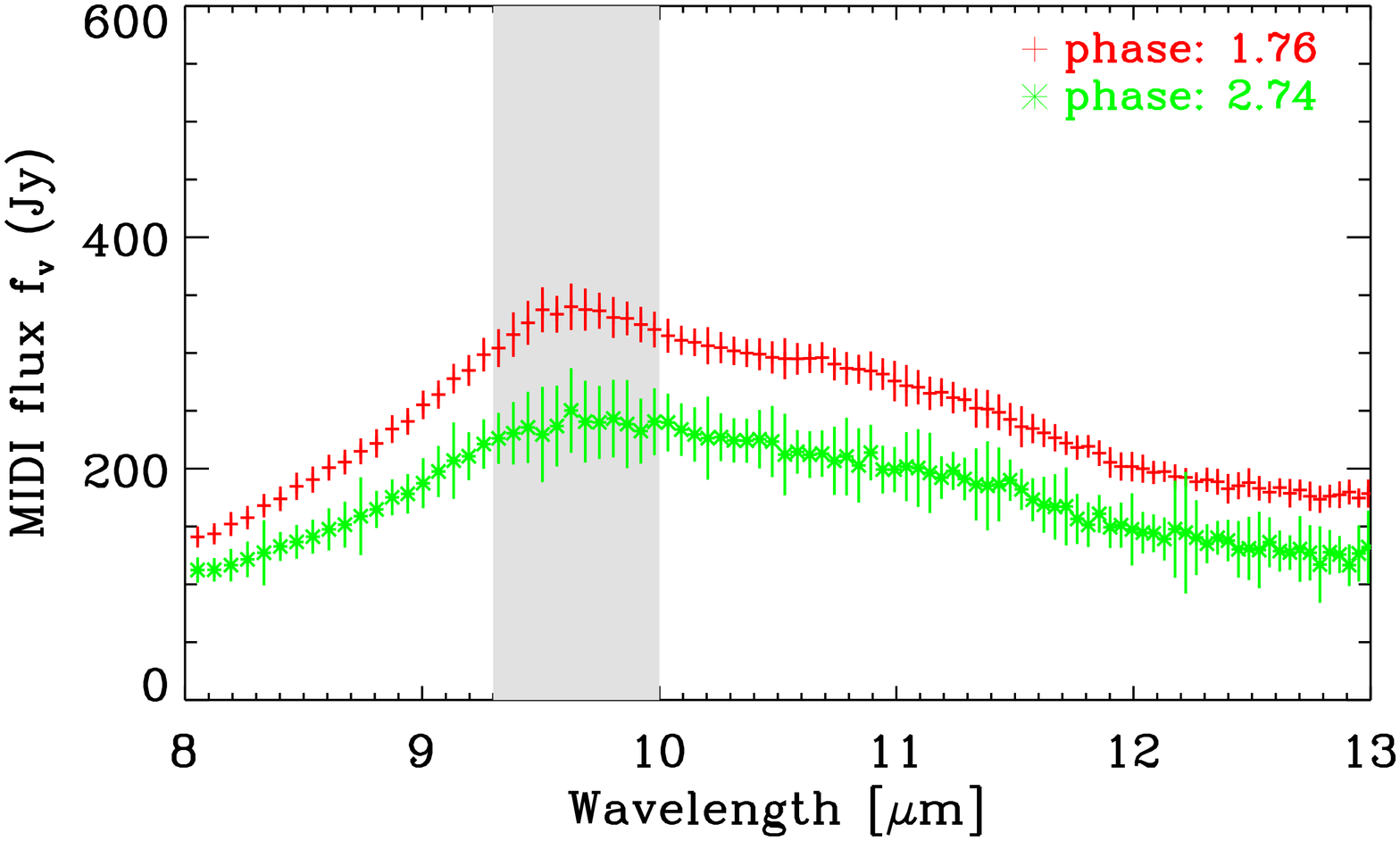}
\includegraphics[height=0.35\textheight,angle=90]{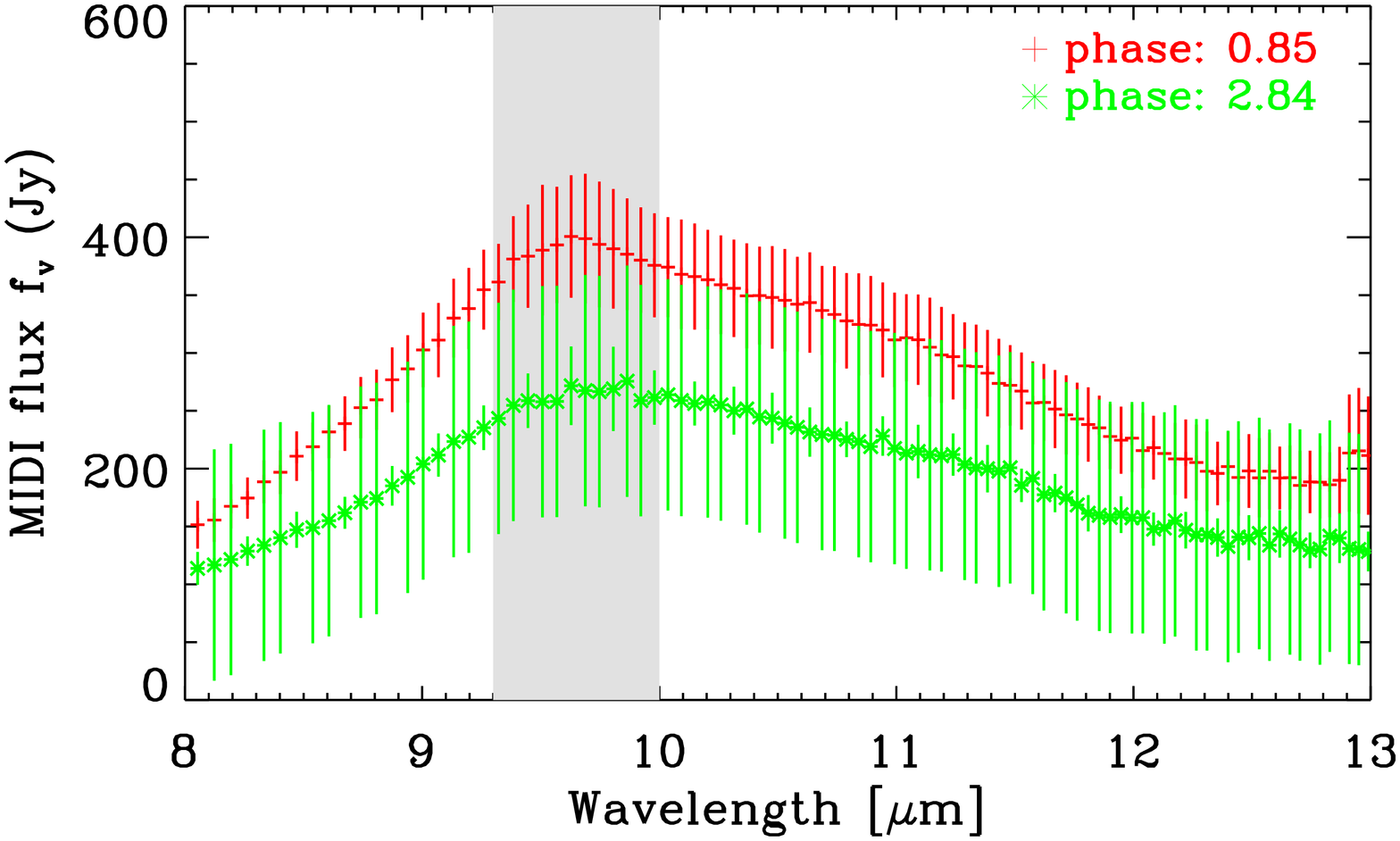}
 \caption{As for Fig.~\protect\ref{phot_ic_1}, but for the same pulsation phase
in consecutive pulsation cycles to investigate cycle-to-cycle
photometry variations. The top panel shows the example of phase $\sim$ 0.75
in cycles 1 and 2. The bottom panel shows the example of phase $\sim$ 0.85
of cycles 0 and 2.}
\label{phot_ctc_2}
\end{figure}

\subsubsection{Deviations from circular symmetry}
Most position angles varied around 40$^\circ$ or 70$^\circ$. Only in
two cases is the P.A. significantly different with values
of 3$^\circ$ and 7$^\circ$. However, in both cases the visibility data
quality is poor. The differential phases are close to zero within
$\sim$\,10-20\,deg, which is close to the calibration
uncertainties that can be reached for MIDI differential phases \citep[e.g.][]{Ohnaka2008}.
Therefore these observations do not allow us to
make any reliable conclusion about the presence or absence
of an asymmetric intensity distribution.\\ 

Summarizing,
our RR~Aql MIDI data do not show any evidence of significant intracycle 
or cycle-to-cycle visibility variations within the examined phase coverage 
of $\sim$ 0.45--0.85 in four consecutive cycles (see Table~\ref{observations}
for details of the phase coverage) and within our visibility accuracies
of $\sim$ 5--20\%. 
The obtained mid-infrared interferometric observations imply either that 
the mid-infrared sizes of molecular and dusty layers of RR~Aql do not 
significantly vary within our phase coverage or that the conducted 
observations are not sensitive enough to detect such 
variations. In addition, the good agreement of visibility data
obtained at the same projected baseline lengths and position angles
for several years
also confirms the good data quality and credibility of the data 
reduction procedure.

\subsection{\textit{N}-band flux monitoring}
\label{sec:fluxmonitoring}
To investigate the intracycle and cycle-to-cycle variability 
of our 
8--13\,$\mu$m flux spectra, we combined the data into groups
of data obtained at similar pulsation 
phases ($\Phi$$_\mathrm{vis}$ $\pm$ 0.15).
The data within each group were averaged. The uncertainty of the averaged
photometry curves was estimated as the standard deviation
of the averaged values.

\subsubsection{Intracycle photometry monitoring}
Figure~\ref{phot_ic_1} 
shows two examples of a comparison of the MIDI calibrated flux spectrum
at different pulsation phases of the same cycle. 
The top panel shows a comparison between phases 0.58 and 0.85 of cycle
0, and the bottom panel shows a comparison between phases 1.53 and 1.76
of cycle 1. Both examples compare data taken at post-maximum and pre-minimum
phases, i.e. with the largest separation in phase that is available in our
dataset.
The RR~Aql flux values within the \textit{N}-band are larger
at the pre-maximum phases than at the post-minimum phases,
corresponding to the light curve of the $V$-magnitude.
The difference is most pronounced toward the silicate emission
feature around 9.8\,$\mu$m and smaller towards the edges
of the MIDI bandpass at 8.0\,$\mu$m and 13.0\,$\mu$m. 
This can most likely be understood as a consequence of 
the silicate absorption
coefficient being largest at short visual wavelengths, so that
the silicate temperature is particularly sensitive to the large
visual flux variations of the central star.
The differences correspond to up to 20\%--35\%, or about
one or two standard deviations.

\subsubsection{Cycle-to-cycle photometry monitoring}
Figure~\ref{phot_ctc_2} 
shows a comparison of 
data observed at the same phase of consecutive pulsation cycles. 
The top panel shows a comparison between phases 1.76 and 2.74, i.e.
phase 0.75 of cycles 1 and 2.
The bottom panel 
shows a comparison between phases 0.85 and 2.84, i.e phase 0.85
of cycles 0 and 2. 
In both examples, the \textit{N}-band fluxes are 
lower in cycle 2 than for the same phase of (a) cycle 1 and (b) 
cycle 0 by up to $\sim$ 100\,Jy, corresponding to about 30\% or
one to two standard deviations. As for the intracycle photometry
variations discussed in the previous paragraph, the differences
are most pronounced toward the silicate emission feature at 
9.8\,$\mu$m.
This might indicate an irregularity in the variability cycle in 
the \textit{N}-band. In the optical band the visual light curve 
clearly shows that the pulsation cycles are not perfectly 
symmetric (see Fig.~\ref{LightCurve}). \\

Summarizing, our data exhibit a 1--2$\sigma$ signature
of intracycle, as well as cycle-to-cycle, flux variations at wavelengths 
of 8--13\,$\mu$m, which are most pronounced toward the silicate
emission feature at 9.8\,$\mu$m. This indication is consistent with
observations by \citet{Monnier1998}, who report temporal variations 
in the mid-infrared spectra of late type stars. In particular, stars with a 
strong emission feature around 9.8\,$\mu$m showed evident changes in the
spectral profile. Considerable phase variation of the \textit{N}-band 
spectra can also be found in \citet{Alvarez1998} and \citet{Guha2011}. 

\section{Modeling of our MIDI data}
\label{sec:modeling}
Self-consistent models describing the dynamic atmosphere
of the central source \textit{and}
the dust shell are still very rare.
This is particularly true for oxygen-rich stars. Nevertheless, 
some advance in this domain has already been successfully achieved
\citep{Ireland2006,Hofner2007,Ireland2008,Hofner2008}. Here, we use an approach of an ad hoc radiative transfer modeling 
of the dust shell where the central stellar source
is described by
readily available and established dust-free 
dynamic model atmosphere series, as
introduced by \citet{Wittkowski2007}.


The dust shell surrounding the central star is modeled by the 
Monte Carlo radiative transfer code mcsim\_mpi \citep{Ohnaka2006}. 
The radiative transfer model requires an assumption on the
spectral energy distribution (SED) of the central stellar source.
Since we expect the photosphere and molecular layers of RR~Aql
to be partly resolved with our MIDI baselines ($V^2$ $\sim$ 0.4 for $\Theta=10$\,mas, $B=120$\,m, $\lambda=10\,\mu$m), we also need to 
model the intensity distribution across the atmospheric layers.
For these purposes, we use the dust-free dynamic model atmospheres, based on 
self-excited pulsation models (P and M series), by 
\citet[][ and references therein]{Ireland2004a,Ireland2004b} as the currently
best available option.

The P and M series were constructed to reproduce the M-type Mira
prototypes $o$ Cet and R Leo. The hypothetical nonpulsating ``parent``
stars have masses $M/M_\odot=1.0$ (P series), and 1.2 (M series),
Rosseland radius $R_p/R_\odot=241$ (P) and 260 (M), and
effective temperatures $T_\mathrm{eff}=2860$\,K (P) and 2750 K (M).
The models have solar abundances, luminosity $L/L_\odot=3470$,
and a pulsation period of 332 days, close to the periods of o~Cet (332 days)
and R~Leo (310 days). They are available for 20 phases in three cycles for
the M series, and 25 phases in four cycles for the P series.
These models have been shown to be consistent with broad-band 
near-infrared interferometric data of o~Cet by
\citet{Woodruff2004} and of R~Leo by \citet{Fedele2005} respectively.
They have also been shown to be consistent with spectrally
resolved, near-infrared interferometric data of the longer period
($\sim$ 430 days) Mira variable S Ori exhibiting a significant variation
in the angular size as a function of wavelength, which was interpreted
as the effect of molecular layers lying above the continuum-forming
layers as predicted by these model series \citep{Wittkowski2008}.
\citet{Tej2003} indicate that the spectral shape of Mira variables
can be reproduced reasonably well even for stars with different 
pulsation periods.
In comparison with the stellar parameters of the P and M model series,
RR~Aql pulsates with a longer period of 394 days. The main sequence 
precursor mass of RR~Aql is
1.2\,$\pm$\,0.2 $M/M_\odot$ \citep{Wyatt1983}, and its spectral type
is M6e-M9 \citep{Samus2004}, versus M5-M9 (P series) and M6-M9.5 (M series).
For the purpose of the modeling effort conducted here,
we can assume that the general properties of the P and M model 
series are also valid for a longer
period variable such as RR~Aql. However, an exact match of 
model parameters and observed parameters as a function of 
phase may not be expected.

The flux and visibility values of the combined atmosphere and dust 
shell (``global``) model are computed for each spectral channel as
\begin{eqnarray}
 f_\mathrm{total} & = & f_\mathrm{star}^\mathrm{att} + f_\mathrm{dust}\\
 V_\mathrm{total} & = & \frac{f_\mathrm{star}^\mathrm{att}}{f_\mathrm{total}} V_\mathrm{star} + \frac{f_\mathrm{dust}}{f_\mathrm{total}} V_\mathrm{dust}\label{e2}\
\end{eqnarray}
where $f_\mathrm{star}^\mathrm{att}$ represents the attenuated flux 
from the dust-free model 
atmosphere\footnote{Note that $f_\mathrm{star}^\mathrm{att}$ may slightly vary across the stellar disk, which is not taken into account here.} (extended by a blackbody approximation for
wavelengths beyond 23\,$\mu$m), $f_\mathrm{dust}$ represents the flux from the dust 
shell alone, $f_\mathrm{total}$ is the addition of these 
two terms, and $V_\mathrm{star}$ and $V_\mathrm{dust}$ are the 
synthetic visibilities computed for the dust-free model 
atmosphere and for the dust shell, respectively.

\begin{table}
\caption{Best-fitting model parameters for each epoch}
\centering 
\begin{tabular}{ l r r r r r r}
\hline\hline 
Epoch & $\Phi_\mathrm{vis}$ & $\tau_V$  & $R_\mathrm{in}/R_\mathrm{Phot}$ &
$p$     & $\Theta_\mathrm{Phot}$ \\
    &            & (sil.) &      (sil.)        &    (sil.)  &[mas]               \\\hline
A    & 0.58 &  3.0 $\pm$ 0.9 &4.5 $\pm$ 1.1 &2.5 $\pm$ 0.4 &6.6 $\pm$ 2.4    \\ 
B    & 0.81 &  3.0 $\pm$ 0.7 &4.5 $\pm$ 1.6 &2.5 $\pm$ 0.5 &7.1 $\pm$ 2.9      \\
C    & 0.86 &  2.5 $\pm$ 0.8 &5.0 $\pm$ 1.3 &2.5 $\pm$ 0.5 &8.0 $\pm$ 2.8     \\
D    & 1.53 &  3.7 $\pm$ 0.6 &3.2 $\pm$ 1.1 &2.5 $\pm$ 0.5 &6.7 $\pm$ 2.6     \\
E    & 1.77 &  2.0 $\pm$ 0.8 &4.5 $\pm$ 1.4 &3.0 $\pm$ 0.4 &8.0 $\pm$ 2.9     \\
F    & 2.46 &  2.7 $\pm$ 0.8 &4.6 $\pm$ 1.1 &3.0 $\pm$ 0.5 &7.7 $\pm$ 2.4      \\
G    & 2.55 &  3.0 $\pm$ 0.6 &3.4 $\pm$ 1.7 &2.1 $\pm$ 0.6 &7.2 $\pm$ 2.5      \\
H    & 2.62 &  2.0 $\pm$ 0.8 &4.5 $\pm$ 1.1 &2.4 $\pm$ 0.5 &8.5 $\pm$ 2.7     \\
I    & 2.74 &  2.0 $\pm$ 0.9 &4.0 $\pm$ 1.5 &2.6 $\pm$ 0.5 &8.5 $\pm$ 2.8      \\
J    & 2.84 &  3.2 $\pm$ 0.7 &3.0 $\pm$ 1.5 &2.0 $\pm$ 0.6 &7.2 $\pm$ 2.6     \\
K    & 3.43 &  5.0 $\pm$ 0.8 &3.0 $\pm$ 1.3 &3.0 $\pm$ 0.5 &7.3 $\pm$ 2.2     \\
L    & 3.54 &  2.3 $\pm$ 0.9 &5.0 $\pm$ 1.5 &2.5 $\pm$ 0.4 &8.2 $\pm$ 2.4     \\
M    & 3.57 &  2.5 $\pm$ 0.7 &4.5 $\pm$ 1.1 &2.5 $\pm$ 0.4 &7.5 $\pm$ 2.4    \\\hline
\end{tabular}
\label{tab:par}
\end{table}

\begin{table}
\caption{Average model parameters}
\centering
\begin{tabular}{ l r r r r r}
\hline\hline 
Model & $\tau_V$  & $\tau_V$ & $R_\mathrm{in}/R_\mathrm{Phot}$ & $p$  & $\Theta_\mathrm{Phot}$ \\
      & (Al$_2$O$_3$)  & (silicate) &  (silicate)                        & (silicate)  &[mas]         \\\hline
 M21n & 0.0     & 2.8 $\pm$ 0.8 & 4.1 $\pm$ 0.7 & 2.6 $\pm$ 0.3 &  7.6 $\pm$ 0.6     \\\hline
\end{tabular}
\label{tab:par_avg}
\end{table}

\subsection{MIDI model parameters}
\label{sec:parameters}
A clear silicate feature is identified in the spectra of RR~Aql
(see Fig.~\ref{iso_iras} and Sect.~\ref{sec:midiresults}).
RR~Aql is one of 31 oxygen-rich stars studied by
\citet{Lorenz-Martins2000}. The authors modeled the stars
using Al$_2$O$_3$ and silicate grains and suggest that the dust chemistry
of RR~Aql contains only silicate grains. In this study we examined both
of these two dust species, Al$_2$O$_3$ grains \citep{Begemann1997,Koike1995} and silicates grains \citep{Ossenkopf1992}.
The amount of dust is described by the optical depths at
$\lambda_0=0.55$\,$\mu$m. The grain size was set to 0.1\,$\mu$m for
all grains. We set the photospheric radius to the well-defined
continuum photospheric radius at $\lambda=1.04\,\mu$m
($R_\mathrm{Phot}=R_\mathrm{1.04}$).
The density gradient was defined by a single power
law $\rho(r) \propto r^{-p}$ with index $p$. The shell thickness was set
to $R_\mathrm{out}/R_\mathrm{in}=1000$.

The global model includes seven parameters. The radiative transfer model 
describing the circumstellar dust shell includes six parameters:
the optical depths $\tau_V$ (Al$_2$O$_3$) and $\tau_V$(silicate), 
the inner boundary radii $R_\mathrm{in}/R_\mathrm{Phot}$ (Al$_2$O$_3$) and
$R_\mathrm{in}/R_\mathrm{Phot}$ (silicate), and the density gradients 
$p_\mathrm{A}$ (Al$_2$O$_3$) and $p_\mathrm{B}$ (silicate). 
The model of the P/M atmosphere series is another parameter.
Here, we used ten models covering one complete cycle 
of the M series: M16n (model visual phase $\Phi_\mathrm{model}=0.60$), 
M18 (0.75), M18n (0.84), M19n (0.90), M20 (0.05), M21n (0.10), M22 (0.25), 
M23n (0.30), M24n (0.40), and M25n (0.50).

We computed a grid of dust-shell models for each of these M models, including all combinations of 
optical depths $\tau_V$ (Al$_2$O$_3$) = 0.0, 0.1, 0.2, 0.5, 0.8; 
$\tau_V$ (silicate) = 2.0, 2.5, 3.0, 3.5, 4.0, 4.5, 5.0; 
$R_\mathrm{in}/R_\mathrm{Phot}$ (Al$_2$O$_3$) = 2.0, 2.5, 3.0; 
$R_\mathrm{in}/R_\mathrm{Phot}$ (silicate) = 2.5, 3.5, 4.5, 5.5, 6.5; 
$p_\mathrm{A}$ (Al$_2$O$_3$) = 2.0, 2.5, 3.0, 3.5; 
and $p_\mathrm{B}$ (silicate) = 2.0, 2.5, 3.0, 3.5.

In a first selection of suitable models, we compared 
the MIDI data of each epoch to our entire grid of models. 
The angular diameter $\Theta_\mathrm{Phot}$ was the only free parameter. 
We increased the uncertainty of the part of the photometric spectra around 9.5\,$\mu$m, which 
is strongly affected by telluric absorption. Otherwise, the weight of each 
data point was given by the corresponding uncertainty. For each epoch we 
kept the best $\sim$30 models with lowest $\chi^2$ values. In a next step, 
a grid with finer steps around these parameters was computed. This procedure 
was repeated several times. The results of the automatic selection were 
visually inspected. We completed the selection with ten models for each 
epoch that agrees with the data best. 


\subsection{MIDI model results}
\label{sec:results}
For each of the 13 epochs we found the best-fitting model parameters that
include the best-fitting set of dust parameters of the radiative transfer model, 
together with the
best-fitting model of the P/M series. We used the procedure outlined 
in Sect.~\ref{sec:parameters}.

Following \citet{Lorenz-Martins2000}, we investigated dust shells
including Al$_2$O$_3$ and/or silicate shells with different inner radii
and density gradients. We obtained best-fit results with a silicate shell
alone, and the addition of an Al$_2$O$_3$ shell
did not result in any improvement in the model fits. 
This result is consistent with \citet{Lorenz-Martins2000}, who 
classify RR~Aql as a source that can be described with a silicate
dust shell alone.

The best-fitting parameters for each epoch are listed in 
Table~\ref{tab:par}. The table lists the epoch, the phase at the epoch, 
the optical depth $\tau_V$, the inner boundary radius 
$R_\mathrm{in}/R_\mathrm{Phot}$, the density distribution 
$p$, and the continuum photospheric angular diameter $\Theta_\mathrm{Phot}$. 
Here, the dust shell parameters are those of the silicate dust shell. The errors of the dust shell parameters are derived in the same way as the standard 
deviation based on the find best-fitting models (Table~\ref{tab:par}).
Table~\ref{tab:par_avg} lists the average model parameters of our
different epochs. The average phase of our observations is                  
$\overline{\Phi_V}=0.64\pm0.15$.

The best-fitting model atmosphere was M21n ($T_\mathrm{model}=2550$\,K,     
$\Phi_\mathrm{Model}$=0.1) for all epochs covering minimum to pre-maximum pulsation phases (0.45--0.85).
The difference between the average phase
of our observation and the phase of the best-fitting model atmosphere
can most likely be explained by the different stellar parameters
of RR\,Aql compared to those of the M series. It means
that $T_\mathrm{eff}$ of RR~Aql at the time of observation is 
similar to $T_\mathrm{eff}$ of model M21n, which has a different
phase, but
its basic parameters are also not exactly those of RR~Aql.

Figure~\ref{A} includes
the model flux and model visibility compared to the observed values for the example of epoch A.
For most epochs, the agreement between the models and the 
observed data is very good, in particular for the visibility spectra. 
The visibility spectra contain information about the radial structure 
of the atmospheric molecular layers and the surrounding dust shells. 
In the wavelength range from 8\,$\mu$m to $\sim$\,9\,$\mu$m, the dust is 
fully resolved with baselines longer than $\sim$\,20\,m, 
and the visibility contribution of 
the dust reaches into 
the second lobe of the visibility function. RR~Aql is characterized by a 
partially resolved stellar disk that includes atmospheric layers, with a 
typical drop in the visibility function $\sim$\,10\,$\mu$m, where the 
flux contribution of the silicate emission is highest and
the flux contribution of the star relative to the total flux 
decreases. Beyond $\sim$\,10\,$\mu$m, the spatially resolved radiation 
from the optically thin dust shell starts to be a notable part of the 
observed total flux. From $\sim$ 10\,$\mu$m to 13\,$\mu$m, the dust contribution 
becomes nearly constant while the stellar contribution increases, and this results in a rebound of the visibility function. 
Photometry spectra also fit well. For some epochs, the models predict 
higher fluxes than observed near the silicate emission feature 
at $\sim$9.8\,$\mu$m.
Generally, our attempt at a  radiative transfer model of the
circumstellar dust shell that uses dynamical model atmosphere series to describe
the central stellar source can reproduce the shape of both 
the visibility and the photometry spectra very well. Our model also follows
the shape of the SED in the range of 
1--40\,$\mu$m (Fig.~\ref{iso_iras}).

The obtained model parameters for the different epochs do not indicate
any significant dependence on phase or cycle. This is consistent
with the result from Sect.~\ref{sec:vismonitoring} that a direct
comparison of visibility values of different phases or cycles did
not show any variability within our uncertainties and that the 
photometric values indicated only small variations 
of up to $\sim$\,2\,$\sigma$.
However, we cannot confirm or deny an eventual phase dependence of 
the dust formation process that affects the 8--13\,$\mu$m visibility and 
photometry values by less than our uncertainties of about 
5--20\% and 10--50\%, respectively.

A silicate dust shell alone, i.e. 
without the addition of an Al$_2$O$_3$ dust shell, provides the
best agreement with our data. The average optical depth of the
silicate dust is $\tau_V$(silicate)=2.8$\pm$0.8 at $\lambda=0.55$\,$\mu$m
(corresponding to 0.03 at $\lambda=8$\,$\mu$m, 0.06 at 
$\lambda=12$\,$\mu$m, and a maximum within 8--12\,$\mu$m
of 0.22 at $\lambda=9.8$\,$\mu$m). The inner radius of the dust
shell expressed in $R_\mathrm{Phot}$ at $\lambda$ = 1.04\,$\mu$m is $R_\mathrm{in}$=4.1$\pm$0.7\,$R_\mathrm{Phot}$. The intensity profile of the dust-free dynamic model atmosphere extends to about 1.5--2 photosphere radii at 10\,$\mu$m \citep[cf.][]{Wittkowski2007}. The 
power-law index of the density distribution is $p$=2.6$\pm$0.3. 
Wind models predict a power-law index of the density distribution of 2 at radii outward the dust formation zone. However, for radii close to the dust formation zone (r$<\sim$10 $R_{\star}$), a larger index is expected \citep[cf, e.g., Fig. 6 in ][]{Wittkowski2007}, which is consistent with our result.

The average photospheric angular diameter results in 
$\theta_\mathrm{Phot}=7.6\pm0.6$\,mas. This value is lower
than the K-band ($\lambda=2.2$\,$\mu$m, $\Delta\lambda=0.4$\,$\mu$m)
UD diameter of $\Theta_\mathrm{UD}=10.73\pm0.66$ mas derived by
\citet{vanBelle2002} at a minimum phase of 0.48. 
This can most likely be explained by the different radius definitions.
Our diameter is a photospheric angular diameter that has already 
been corrected for effects of molecular layers lying above the 
continuum photosphere using the prediction by the best-fitting 
model atmosphere,
while the diameter by van Belle et al. is a broad-band uniform
disk diameter that still contains contamination by molecular layers.
It is known that broad-band UD diameters may widely overestimate the 
photospheric radius owing to contamination by molecular layers lying above 
the photosphere \citep[cf., e.g.,][]{Ireland2004a,Fedele2005,Wittkowski2007}.
With the parallax of $\pi=1.58\pm0.40$ mas \citep{Vlemmings2007},
our value for $\theta_\mathrm{Phot}$ corresponds to a photospheric
radius of $R_\mathrm{phot}=520^{+230}_{-140}\,R_\odot$.
Together with the bolometric magnitude of RR~Aql of $m_\mathrm{bol}=3.71$
and $\Delta m_\mathrm{bol}=1.17$ from \citet{Whitelock2000},
our value for $\theta_\mathrm{Phot}$ at $\overline{\Phi_V}\,\sim\,0.64$
corresponds to an effective temperature of 
$T_\mathrm{eff}\sim 2420 \pm 200$\,K. 
This value is consistent with the effective temperature of the
best-fitting model atmosphere M21n, which is $T_\mathrm{eff}=2550$\,K.

\begin{table*}
\caption{Model simulations, comparing two models that
differ in one or more parameters (marked
by bold face).}
\label{table:4}
\centering
\begin{tabular}{ l r r r r r r r r r r}
\hline\hline
Simulation &Model & $\Phi_{\mathrm{vis}}$ & $\tau_V$ &
$\tau$$_V$ & $R_\mathrm{in}/R_\mathrm{Phot}$ & $R_\mathrm{in}/R_\mathrm{Phot}$
& $p$ & $p$  & $\Theta_\mathrm{Phot}$ & Projected\\
 &  & &Al($_2$O$_3$)    &(silicate) &(Al$_2$O$_3$) &(silicate)                           &(Al$_2$O$_3$)    &(silicate)       &[mas]  & baseline [m]       \\
 &  & &tva              &tvb        &ra            &rb                                   &pa               &pb               &diam.  &$B_p$               \\\hline
1  & {\bf M21n}& {\bf 0.1} &0.0&3.0&2.2&4.5&2.5&2.5&{\bf 8.5}&60  \\
   & {\bf M23n}& {\bf 0.3} &0.0&3.0&2.2&4.5&2.5&2.5&{\bf 7.0}&60  \\
            \noalign{\smallskip}
2  & {\bf M21n}&{\bf 0.1}&0.0&{\bf 3.0}&2.2&{\bf 4.5}&2.5&2.5&{\bf 8.5}&60  \\
   & {\bf M23n}&{\bf 0.3}&0.0&{\bf 4.0}&2.2&{\bf 3.8}&2.5&2.5&{\bf 7.0}&60  \\
            \noalign{\smallskip}
3a & M21n&0.1&0.0&{\bf 3.0}&2.2&{\bf 4.5}&2.5&2.5&8.5&{\it 60}  \\
   & M21n&0.1&0.0&{\bf 5.0}&2.2&{\bf 3.0}&2.5&2.5&8.5&{\it 60}  \\
            \noalign{\smallskip}
3b & M21n&0.1&0.0&{\bf 3.0}&2.2&{\bf 4.5}&2.5&2.5&8.5&{\it 40}  \\
   & M21n&0.1&0.0&{\bf 5.0}&2.2&{\bf 3.0}&2.5&2.5&8.5&{\it 40}  \\
            \noalign{\smallskip}
3c & M21n&0.1&0.0&{\bf 3.0}&2.2&{\bf 4.5}&2.5&2.5&8.5&{\it 25}  \\
   & M21n&0.1&0.0&{\bf 5.0}&2.2&{\bf 3.0}&2.5&2.5&8.5&{\it 25}  \\
            \noalign{\smallskip}
3d & M21n&0.1&0.0&{\bf 3.0}&2.2&{\bf 4.5}&2.5&2.5&8.5&{\it 15}  \\ 
   & M21n&0.1&0.0&{\bf 5.0}&2.2&{\bf 3.0}&2.5&2.5&8.5&{\it 15}  \\
            \noalign{\smallskip}
4  & M21n&0.1&{\bf 0.0}&3.0&2.2&4.5&2.5&2.5&8.5&60  \\
   & M21n&0.1&{\bf 0.6}&3.0&2.2&4.5&2.5&2.5&8.5&60  \\
\hline
\end{tabular}
\tablefoot{Here, $\Phi_{\mathrm{vis}}$ is the phase, $\tau_V$ the optical depths, $R_\mathrm{in}/R_\mathrm{Phot}$ the inner boundary radii, $p$ the density gradients, $\Theta_\mathrm{Phot}$ the angular diameter.}
\label{experiments}
\end{table*}

\begin{figure}
\includegraphics[height=0.22\textheight]{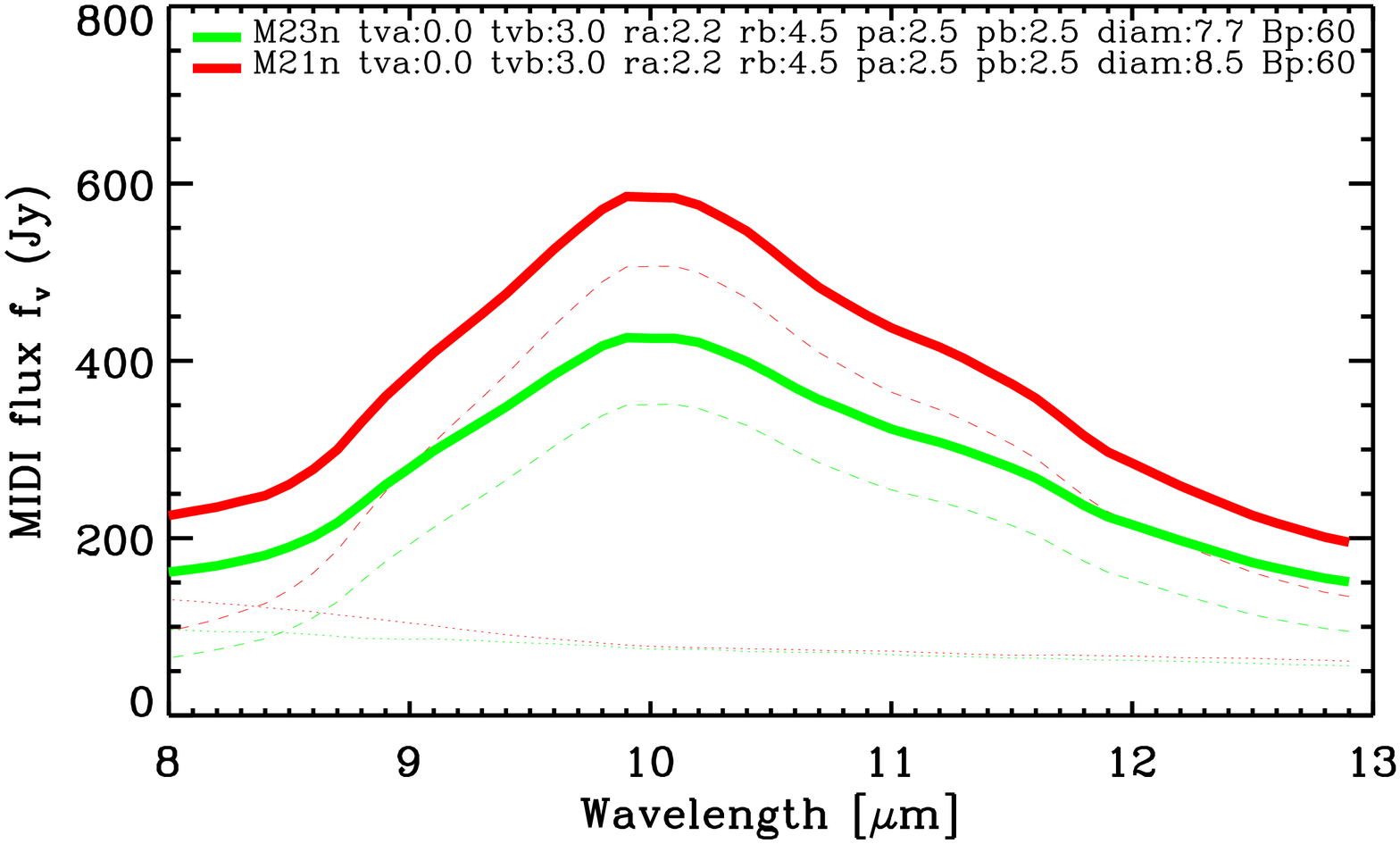}
\includegraphics[height=0.22\textheight]{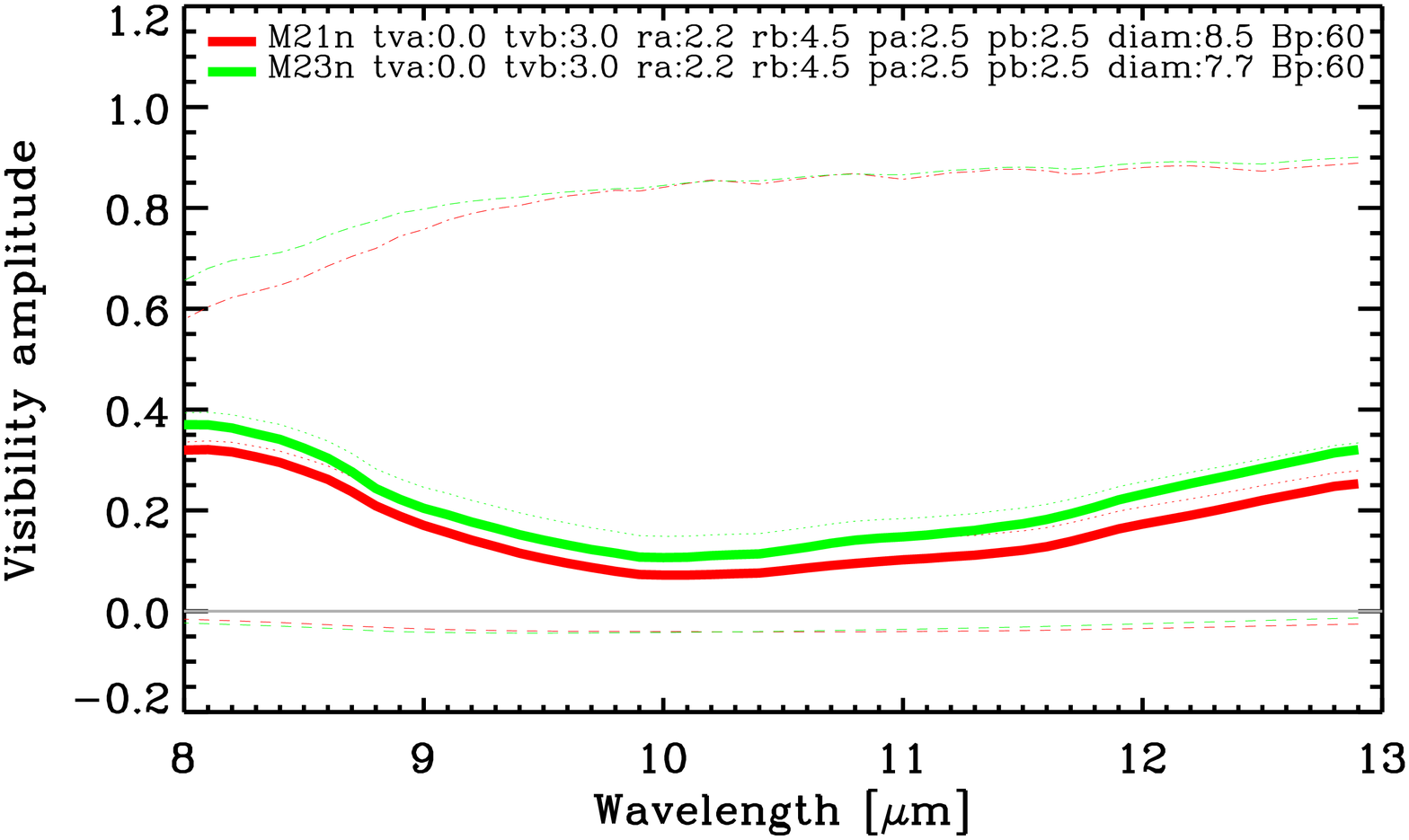}
 \caption{Simulation 1. Synthetic flux (top)
and visibility amplitude (bottom) in the wavelength range 8--13\,$\mu$m.
The solid
lines represent the global models, the dashed-dotted lines denote
the unattenuated stellar contribution i.e. 
$V_\mathrm{star}$ in Eq.~\ref{e2}, the dotted lines denote
the attenuated stellar contribution, and the dashed lines denote
the dust shell contributions.
This simulation compares two models consisting of the same dust shell
parameters but different model atmospheres, the post-maximum
model atmosphere M21n ($\Phi_\mathrm{model}=0.10$), and the
minimum model atmosphere M23n ($\Phi_\mathrm{model}=0.30$).
The photospheric angular diameter is assumed to be larger
at post-maximum phase than at minimum phase (8.5 mas compared to
7 mas). The projected baseline length is 60\,m.
For the exact model parameters, see Table~\protect\ref{experiments}.
They describe variations around our best-fitting model for RR~Aql.}
\label{test1}
\end{figure}

\begin{figure}
 \includegraphics[height=0.22\textheight]{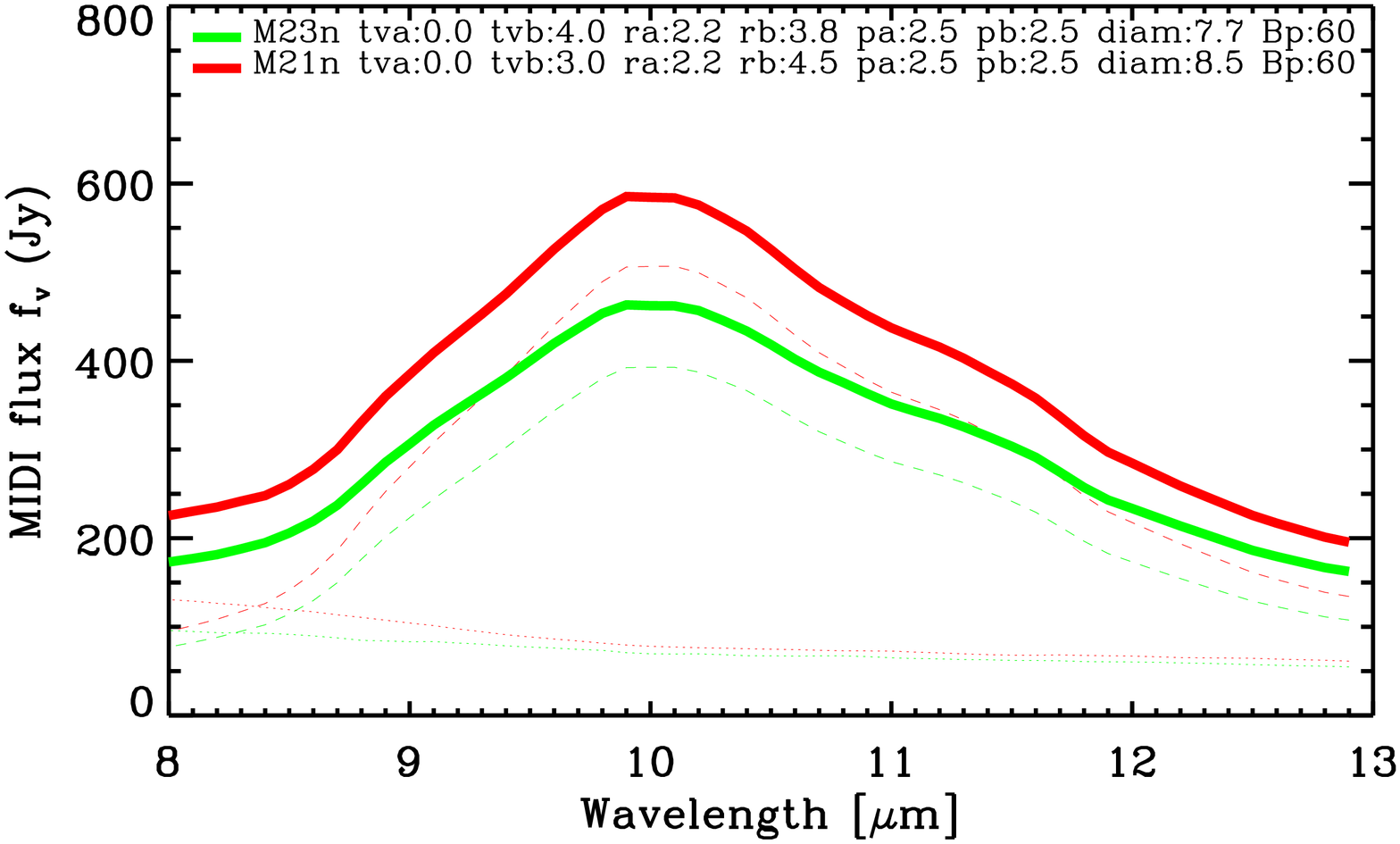}
 \includegraphics[height=0.22\textheight]{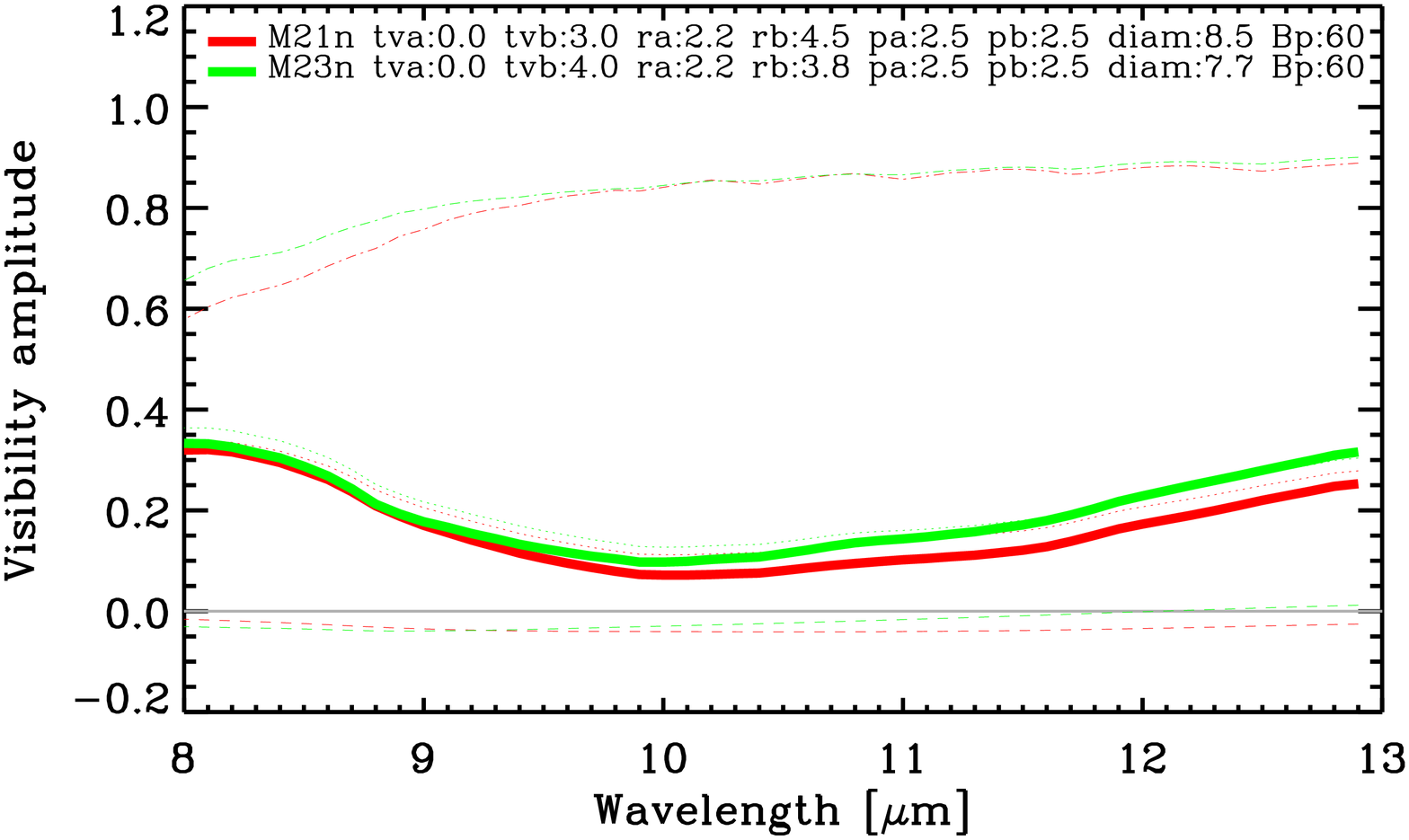}
 \caption{Simulation 2. Like simulation 1 in Fig.~\ref{test1},
but for the parameters of simulation 2 comparing 
a post-maximum atmosphere model and a minimum model as in 
simulation 1, but where
the dust is also assumed to be closer to the star with larger optical
depth at minimum phase and farther from the star with lower optical
depth at post-maximum phase.}
\label{test2}
\end{figure}

\begin{figure}
\includegraphics[height=0.22\textheight]{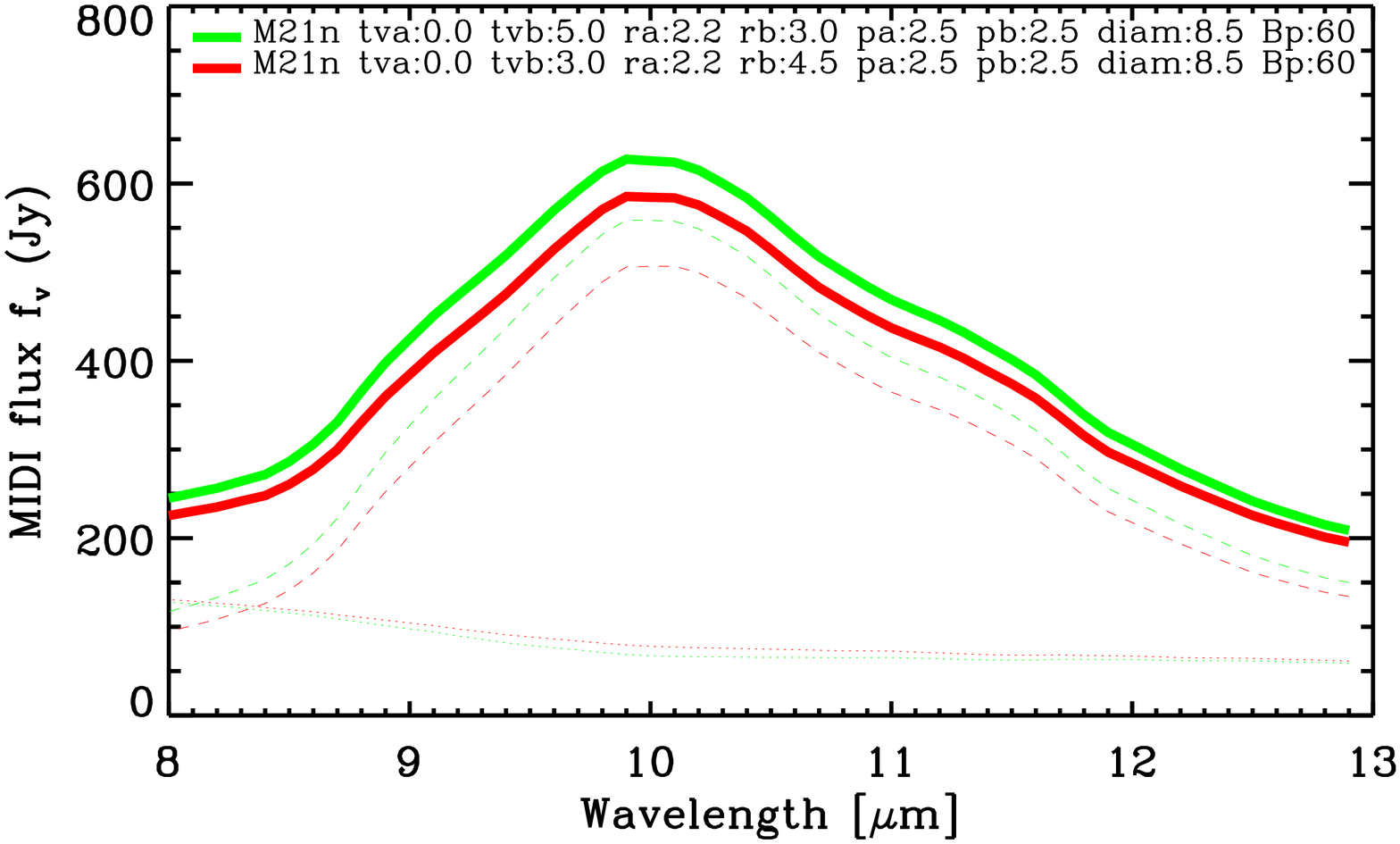}
\includegraphics[height=0.22\textheight]{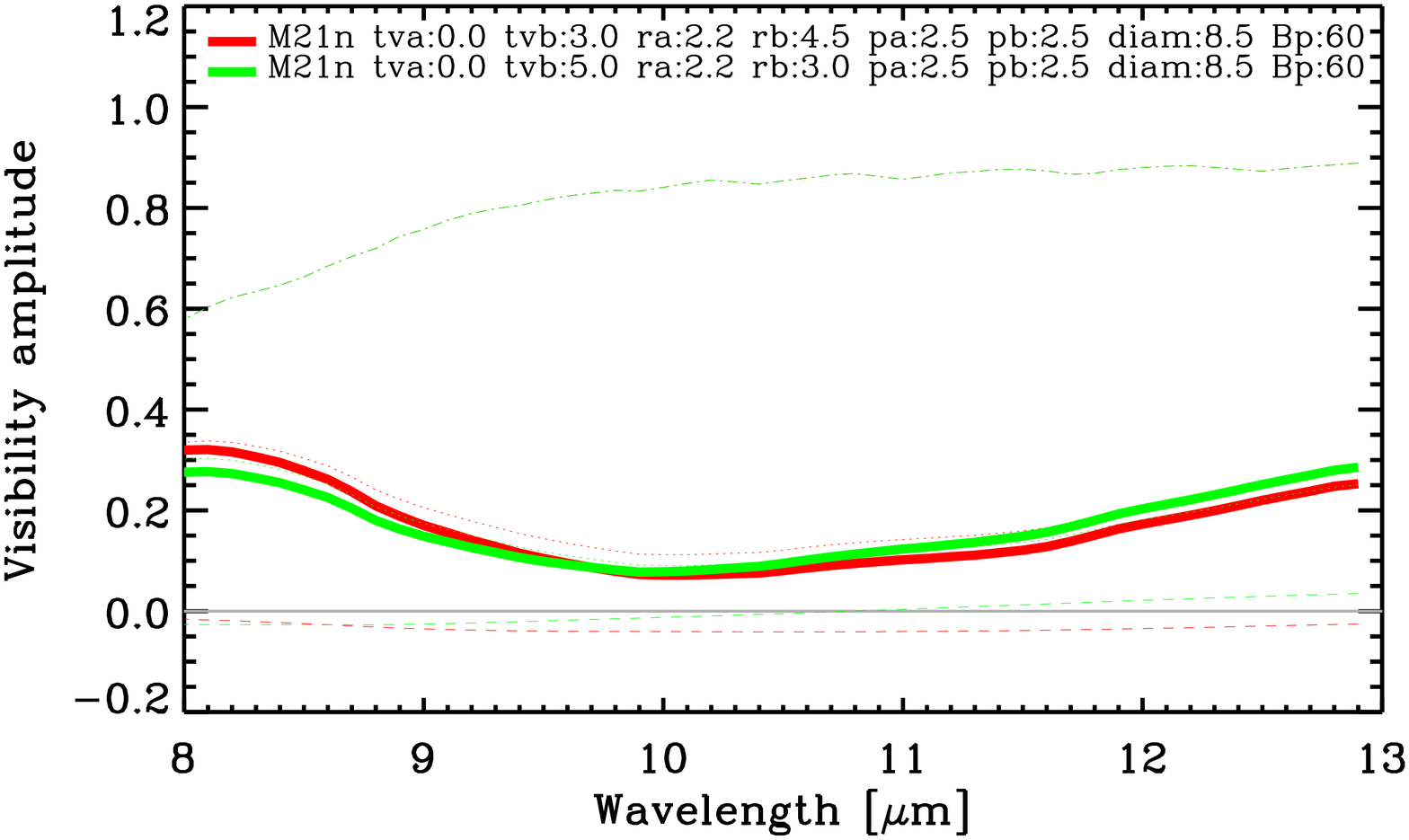}
 \caption{Simulation 3. As for simulation 2 in Fig.~\protect\ref{test2},
but where the atmosphere model is not varied.}
\label{test3}
\end{figure}

\begin{figure}
\includegraphics[height=0.22\textheight]{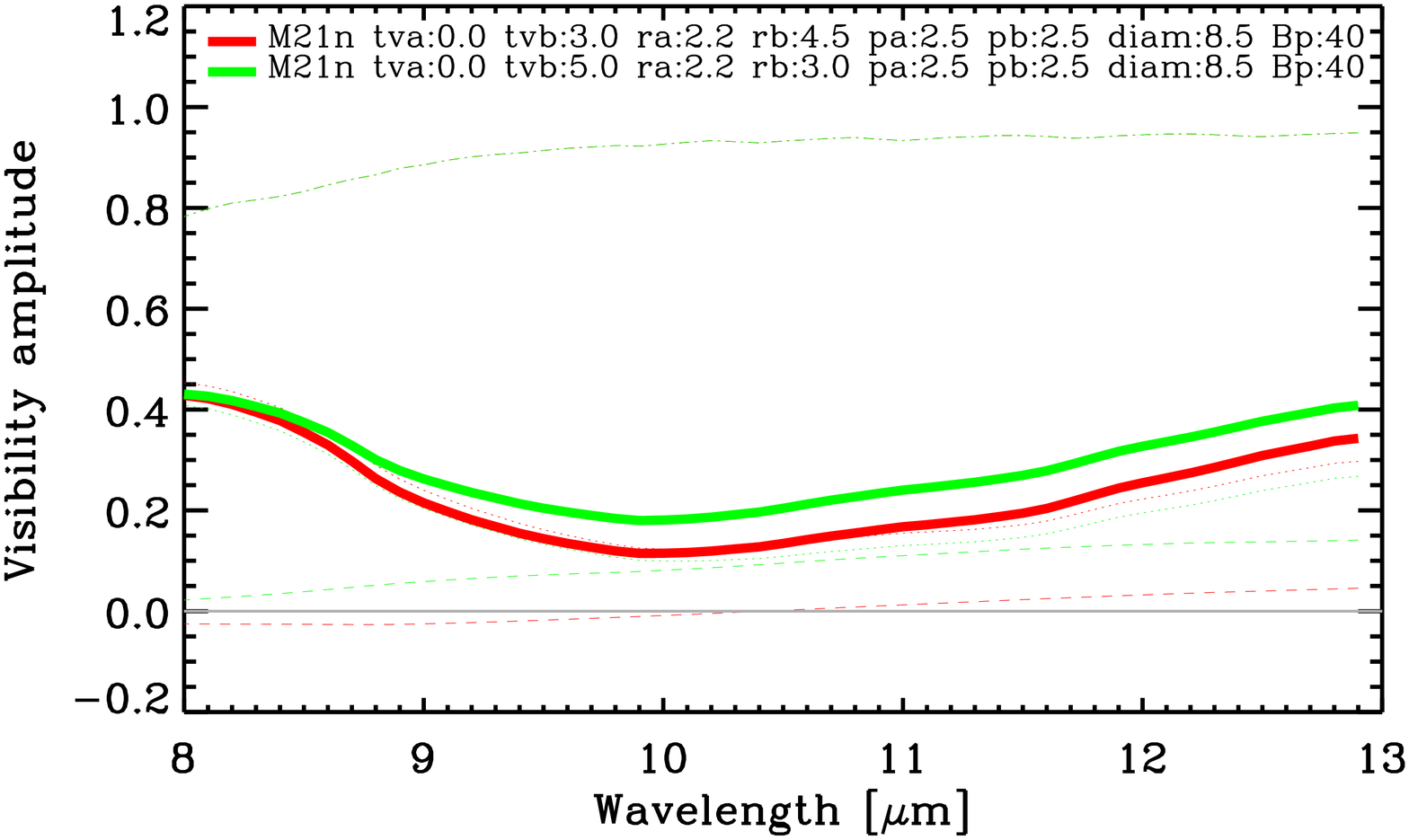}
\includegraphics[height=0.22\textheight]{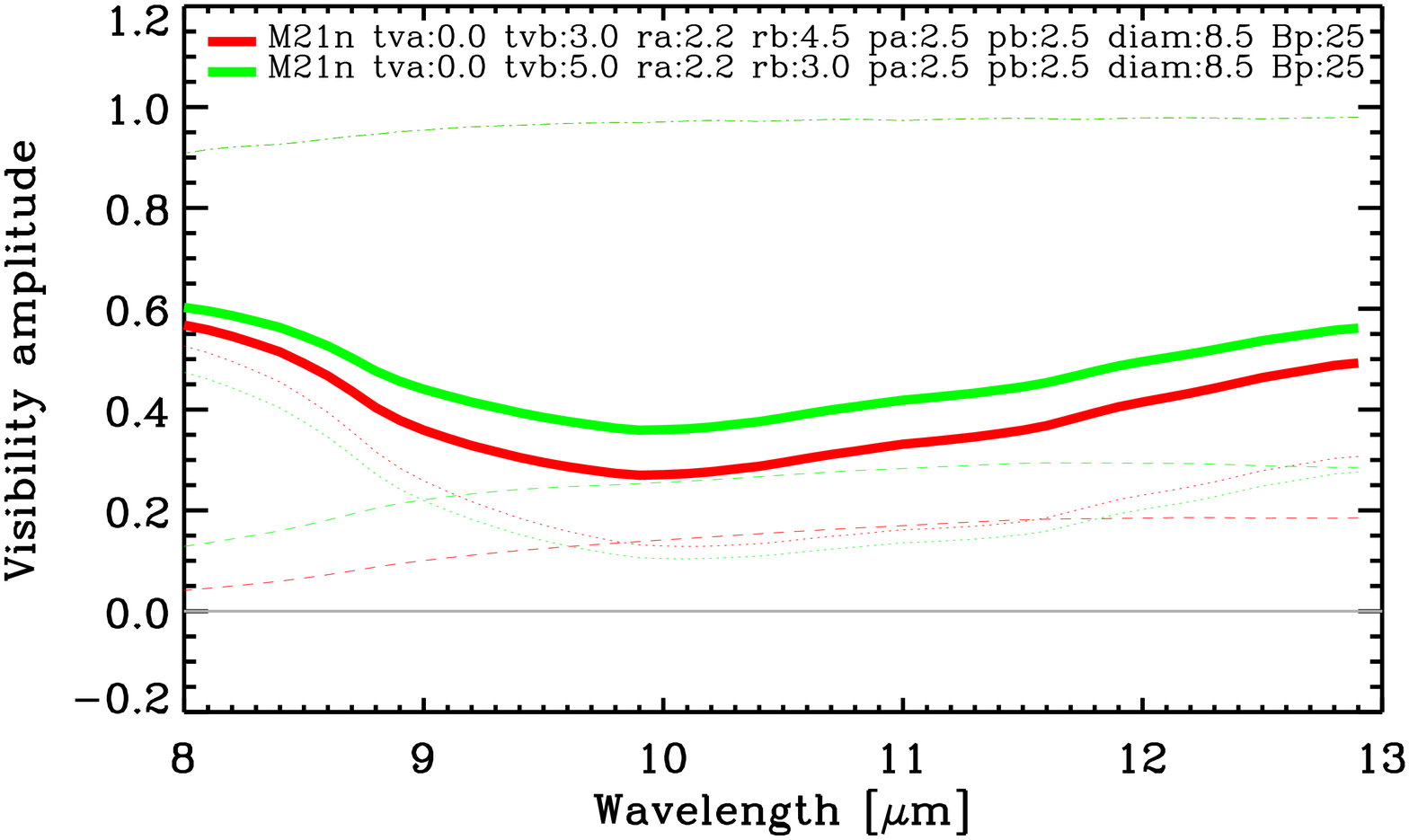}
\includegraphics[height=0.22\textheight]{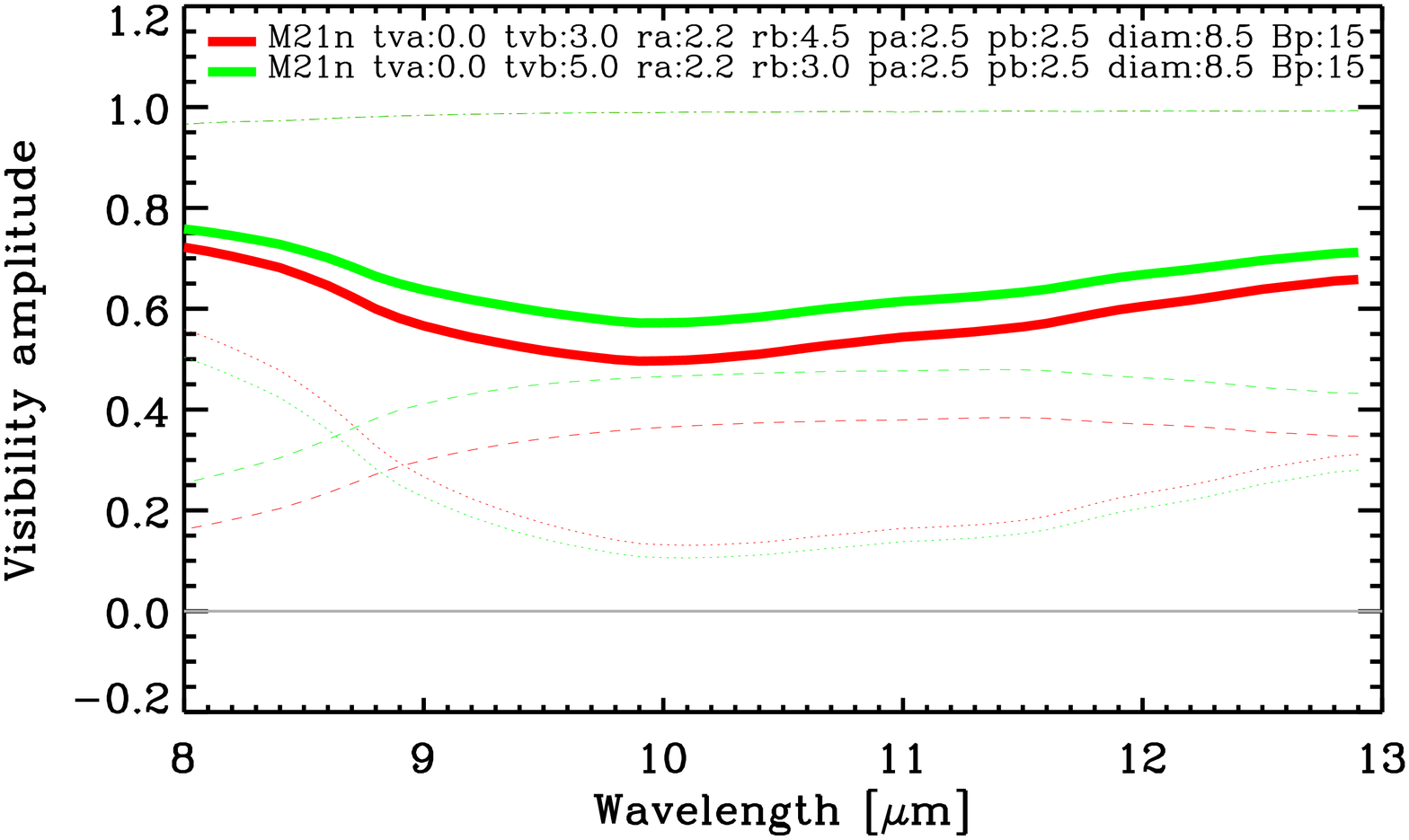}
\caption{Simulation 3. As Fig.~\protect\ref{test3},
but showing the visibility spectra for different projected 
baseline lengths of (from to bottom) 40\,m, 25\,m, and 15\,m.}
\label{test3b}
\end{figure}

\begin{figure}
\includegraphics[height=0.22\textheight]{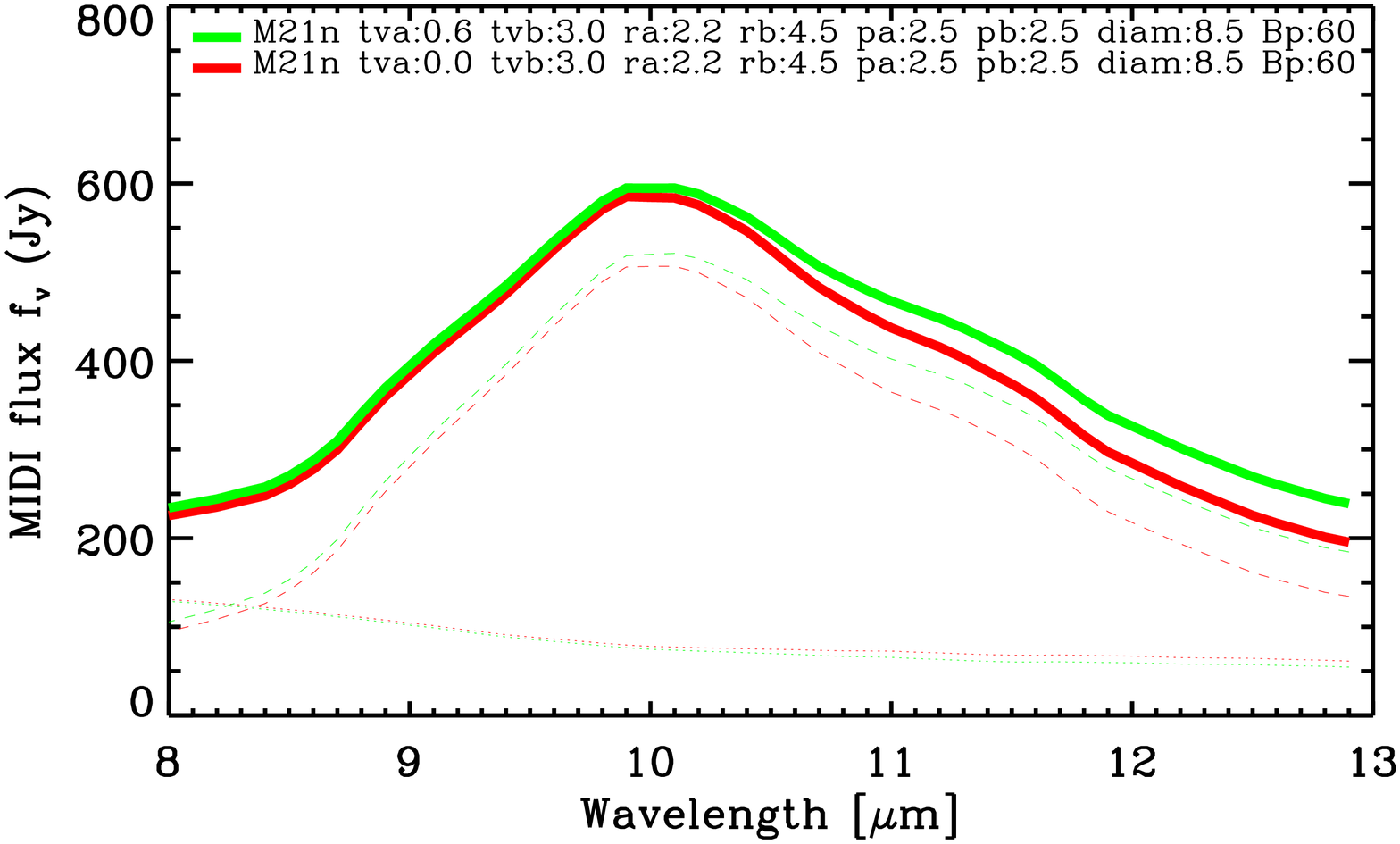}
\includegraphics[height=0.22\textheight]{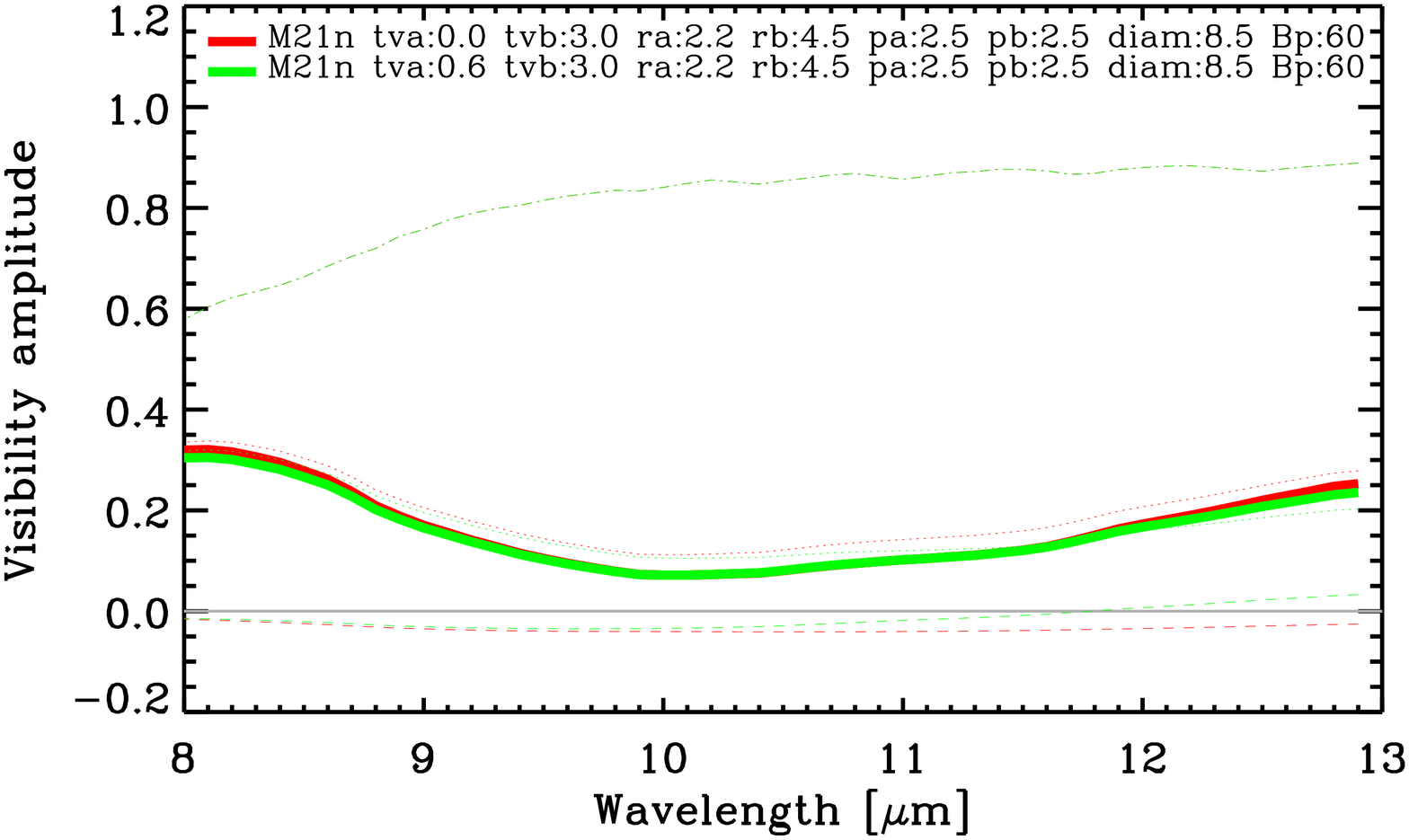}
 \caption{Simulation 4. Comparison of a model with a 
silicate shell only and a model that also includes an 
Al$_2$O$_3$ dust shell with lower optical depth. 
For the model parameters, see Table~\protect\ref{experiments};
For the description of the figure, see the caption of
Fig.~\protect\ref{test1}.}
\label{test4}
\end{figure}


\subsection{Model simulations}
\label{sec:syntheticspectra}
We performed a number of model simulations in order to investigate 
the visibility and photometry variations that are theoretically expected
in the 8--13\,$\mu$m wavelength range for the typical parameters
of RR~Aql as determined above in Sect.~\ref{sec:results}.
We based our simulations on a global model with typical parameters
similar to those derived for RR~Aql (cf. Table~\ref{tab:par_avg}),
and investigated the effects of expected variations in the atmosphere
model and the dust shell parameters during a pulsation cycle on the 
observable photometry and visibility spectra. We mainly used a mean
projected baseline length of 60\,m. As in Sect.~\ref{sec:modeling},
the global model consists of a radiative transfer models of the Al$_2$O$_3$ and
silicate dust shells where the central source is described by a dust-free dynamic model atmosphere.
Table~\ref{experiments} lists the performed simulations including the 
phase of the model, the optical depth $\tau_V$ (Al$_2$O$_3$) and 
$\tau_V$ (silicate), the inner boundary radii of the dust shell 
$R_\mathrm{in}/R_\mathrm{Phot}$ (Al$_2$O$_3$) and 
$R_\mathrm{in}/R_\mathrm{Phot}$ (silicate), the power-law indices
of the density distributions $p$ (Al$_2$O$_3$) and $p$ (silicate), 
and the continuum photospheric angular diameter $\Theta_\mathrm{Phot}$.  
We assume a phase dependence of the angular photospheric diameter of
about 20\% \citep{Thompson2002a,Ireland2004a}.
Figures~\ref{test1}~-~\ref{test4} show the results from the simulations. 
Each simulation compared two models, where one or more of the model
parameters have been varied.

\paragraph{Simulation 1} (Fig. ~\ref{test1})
In simulation 1, we investigate the effect of model atmospheres at 
different pulsation phases and effective temperatures. We compared model M21n at post-maximum 
pulsation phase ($\Phi_\mathrm{model}=0.10$, $T_\mathrm{eff}=2550$\,K) and model M23n at 
pre-minimum pulsation phase ($\Phi_\mathrm{model}=0.30$, $T_\mathrm{eff}=2230$\,K), applying the 
same dust shell parameters for both models. Here, the phase difference between M21n and M23n of 0.2 corresponds to the phase difference of our observations ($\sim$0.64$\pm$0.2) The dust parameters were 
based on the average parameters (see Tab.~\ref{tab:par_avg}) of RR~Aql
derived in Sect.~\ref{sec:results}. The projected baseline length 
was set to $B_p=60$\,m, which is a typical mean value of our MIDI observations. 
The overall shapes of the visibility spectra 
$V_\mathrm{total}$ (M21n) and $V_\mathrm{total}$ (M23n) are similar.
The differences between the models are wavelength-dependent in the range 
of 5--20\%. The shape of the photometry spectra are similar for both 
tested models $f_\mathrm{total}$ (M21n) and $f_\mathrm{total}$ (M23n). 
The largest difference of $\sim$\,25\% is seen around 10\,$\mu$m, with 
less flux at minimum pulsation phase. 
Compared to the uncertainties of our MIDI measurements of
$\sim$\,5--20\% in the visibility spectra and of $\sim$\,20\% on average in the photometric spectra, simulation 1 is consistent with
our non-detection of visibility variations and a marginal detection
of photometry variations.

\paragraph{Simulation 2} (Fig. ~\ref{test2})
Compared to simulation~1, we adjusted the parameters according to the 
assumption that the outer 
layers of the atmosphere are cooler near minimum visual pulsation phase, 
and therefore more dust grains can form, and a higher molecular opacity 
can be expected. This theoretical assumption is 
consistent with mid-infrared interferometric observations 
\citep{Wittkowski2007}, where the observed data indicated more dust 
formation near minimum pulsation phase, with inner boundary radii 
of the dust shell located closer to the star. We increased the optical 
depth for model M23n. We also set a more extended dust shell after 
visual maximum for model M21n in comparison to the model M23n. This setting 
results in spectra with differences in the range of 2-19\% for the 
visibility values. The photometry differs again mostly around 10\,$\mu$m with a 
maximum difference of 20\%. Compared to simulation 1, the results
are very similar, but with less differences in the photometry spectra. 
As for simulation~1, the results are
consistent with our not detecting visibility variations and 
marginally detecting photometry variations within our measurement
uncertainties.

\paragraph{Simulation 3} (Figs.~\ref{test3} and \ref{test3b})
Here, we investigate the effect of different dust shell parameters
(as in simulation 2), but keeping the M model constant. 
We used model M21n (phase 0.1), which was the best-fitting model
to our RR~Aql epochs and also used a constant photospheric angular
diameter. We compared two (silicate) dust models as in simulation 2,
where in one model the optical depth is lower and the inner radius of the
dust shell larger compared to the other model.
These sets of model parameters lead to very similar results, where the 
visibility spectra almost overlap, with a maximum
difference of 4\% for a projected baseline length of 60\,m 
(Fig.~\ref{test3}). The photometry spectra are also very similar 
with a maximum difference of 7\%. 
Figure~\ref{test3b} shows the visibility results for different 
projected baseline lengths of 40\,m, 25\,m, and 15\,m.
With projected baseline lengths of less than 60\,m, the difference 
between the visibility values become more pronounced (10\%
difference beyond 10\,$\mu$m with projected baseline lengths of 
40\,m and 25\,m). 
This can be understood because the dust shell is over-resolved with
a projected baseline length of 60\,m, so that the visibility cannot be
sensitive to variations in the geometry of the dust shell.
With lower projected baseline length, the visibility values are located
in the first lobe of the visibility function, hence more sensitive to 
variations of the dust shell geometry. 
However, with the even lower projected baseline lengths of 15\,m,
the 
difference between individual models again starts to be less 
pronounced ($<$7\%). This indicates an optimum projected
baseline length of $\sim$\,25--40\,m to characterize the dust
shell geometry for sizes as found for RR~Aql.

\paragraph{Simulation 4} (Fig.~\ref{test4})
Furthermore, we investigate the effect of adding small amounts of 
Al$_2$O$_3$ grains to a thus far pure silicate dust shell. We examined $\tau_V$(Al$_2$O$_3$) of 0.3 and 0.6. Figure~\ref{test4} shows the simulation 
with $\tau_V$(Al$_2$O$_3$) of 0.6, showing that
the addition of small amounts of Al$_2$O$_3$ grains leads to almost 
identical photometry and visibility spectra (differences $<$2\%).


\section{Discussion}
Our MIDI observations of RR~Aql did not show 
significant variations in the 8--13\,$\mu$m visibility
within the examined pulsation phases between 0.45 and 0.85
within a total of four pulsation cycles, and showed only marginal 
variations in the 8--13\,$\mu$m flux. 

We performed model simulations with expected variations of
the pulsation phase of the innermost dust-free atmosphere model 
and of the parameters of the 
surrounding dust shells. These model simulations show that 
visibility variations 
are indeed not expected for the parameters and
observational settings of RR~Aql
at wavelengths of 8--13\,$\mu$m within the 
uncertainties of our observations. Variations in the flux
spectra may in some cases just be detectable. 
Thus, our observational result of a constant visibility
and only slightly varying flux at wavelengths
of 8--13\,$\mu$m are consistent with, and not contradicting,
theoretical expectations of a pulsating atmosphere.

Our model simulations indicate that detections of pulsation effects 
at mid-infrared wavelengths would in particular benefit from smaller 
uncertainties in the photometric spectrum than in our current data. 
Also, a wide range of 
projected baselines at each phase, for RR~Aql in particular including 
baseline lengths around 20-30\,m, would help us to distinguish models 
of different pulsation phases.

For our analysis, we have considered silicate (Ossenkopf et al. 1992)
and Al$_2$O$_3$ \citep{Begemann1997,Koike1995} dust 
species, following the work by \citet{Lorenz-Martins2000}, and 
used fixed grain sizes of 0.1\,$\mu$m. It is known that other 
dust species with a more complex gran size distributions may occur 
in the circumstellar environment of Mira variables \citep[e.g.,][and references therein]{Hofner2008,Molster2003}. We showed that a model 
including only a silicate dust shell can reproduce the observed 
RR~Aql visibility and flux spectrum at 8--13\,$\mu$m. The addition of 
an Al$_2$O$_3$ dust shell with comparable low optical depth did not 
significantly improve the fit to our data. However, our model simulations 
have shown that our 8--13\,$\mu$m visibility and flux values are not
sensitive to the addition of an Al$_2$O$_3$ dust shell with low
optical depth within our uncertainties. As a result, we cannot exclude
the presence of an inner Al$_2$O$_3$ dust shell in addition to the
silicate dust shell. \citet{Woitke2006} uses dynamical models for dust-driven winds of
oxygen-rich AGB stars, including frequency-dependent radiative transfer,
and finds that dust temperatures strongly depend on material.
Two dust layers are formed in his dynamical models,
almost pure glassy Al$_2$O$_3$ close to the star (r $>$∼ 1.5 $R_\mathrm{star}$) and
the more opaque Fe-poor Mg-Fe-silicates farther out at 4--5 $R_\mathrm{star}$. 

The dust-free dynamic model atmospheres predict a significant dependence 
of the characteristics of the molecular layers on the stellar pulsation 
phase at near-infrared wavelengths \citep{Ireland2004a,Ireland2004b}. 
There are a few near-infrared interferometric observations, which 
detected a clear variation in the continuum angular diameters with 
pulsation phase \citep{Thompson2002b,Perrin1999,Young2000b,Woodruff2004}. 
Spectrally resolved near-infrared interferometric measurements
at different phases, 
such as the AMBER observations by \citet{Wittkowski2008} but at more
than one phase, promise to lead to stronger constraints on 
dynamic model atmospheres.

Our model simulations based on the combination of dynamic model
atmospheres with a radiative transfer model of the dust shell
predict only small variations with phase at mid-infrared wavelengths.
The difference near a  wavelength
of 10\,$\mu$m amounts to $\sim$\,5\% for the visibility values and 
to $\sim$\,25\% for the photometry values.
These results lead to the suggestion that the stellar photosphere
and overlying molecular layers pulsate, which is demonstrated by the 
diameter variations in the near-infrared, but that these pulsations
cannot be detected for RR~Aql by our 8--13\,$\mu$m interferometry 
within our uncertainties. In addition, our visibility uncertainties
do not allow us to exclude variations in the geometry and optical
depth of the dust shell as a function of pulsation phase as
observed previously for other targets \citep{Lopez1997}. Possible 
explanations for not detecting a phase-dependence
of the dust shell parameters in our study compared to the ISI
observations include:
(i) longer baseline lengths in our study that often fully
resolve the very extended silicate dust shell;
(ii) our limited phase coverage between minimum and pre-maximum
(0.45--0.85) phases. Possible variations over the whole pulsation
cycle cannot be excluded.

The observed variability of the 8--13\,$\mu$m flux at a significance
level of 1--2$\sigma$ may indicate variations in the stellar radiation
re-emitted by the dust and/or changes in either the dust geometry or optical
depth.


\section{Summary and conclusions}
We have investigated the 
circumstellar dust shell and characteristics of the atmosphere 
of the oxygen-rich Mira variable RR~Aql 
using mid-infrared interferometric observations. 
We observed RR~Aql with the VLTI/MIDI instrument 
at different pulsation phases in order to monitor the photometry 
and visibility spectra. A total of 57 observations were combined into 
13 epochs covering four pulsation cycles between April 2004 and July 2007,
and covering pulsation phases between minimum and pre-maximum phases
(0.45--0.85). 

We modeled the observed data with an ad-hoc radiative transfer model 
of the dust shell using the radiative transfer code mcsim\_mpi by \citet{Ohnaka2006}. 
In this way, we used a series of dust-free 
dynamic model atmospheres based on self-excited pulsation models \citep[M series,][]{Ireland2004a,Ireland2004b} to describe the intensity profile
of the central source.
This study represents the first comparison between interferometric 
observations and theoretical models over an extended range of pulsation 
phases covering several cycles.

Our main observational results 
are as follows 
\begin{itemize}
\item 
The interferometric data do not show any evidence of intracycle visibility variations. \end{itemize}

\begin{itemize}
\item 
The data do not show any evidence of cycle-to-cycle visibility variations. \end{itemize}

\begin{itemize}
\item 
The 8--13\,$\mu$m flux suggests intracycle and cycle-to-cycle photometry variations at a significance level of 1--2$\sigma$. Follow-up observations with higher accuracy using a dedicated photometric instrument, such as VISIR at the VLT, are needed to confirm this result.\end{itemize}

These observational results can be explained by 
dynamic model atmospheres and variations 
in the dust shell parameters.
Simulations 
using different phases of the dynamic model atmosphere
and different sets of dust shell parameters 
predict visibility variations that are lower than or close to the observed visibility uncertainties (5-20\,\%).
Model-predicted variations of the photometry spectra 
are largest around wavelengths of 10\,$\mu$m with
difference of up to $\sim$25\% corresponding to up to 1--2$\sigma$. 

The best-fitting model for our average pulsation phase
of $\overline{\Phi_V}=0.64\pm0.15$ includes 
a silicate dust shell with an optical depth of
$\tau_V (silicate)=2.8\pm0.8$, an inner radius of
$R_\mathrm{in}=4.1\pm0.7$\,$R_\mathrm{Phot}$, and
a power-law index of the density distribution of $p=2.6\pm0.3$.
The corresponding best-fitting atmosphere model of the series 
used to describe the central intensity profile is
M21n ($T_\mathrm{model}=2550$\,K, $\Phi_\mathrm{model}=0.1$)
with a photospheric
angular diameter of $\theta_\mathrm{Phot}=7.6\pm0.6$\,mas.
The photospheric angular diameter corresponds to a photospheric
radius of $R_\mathrm{phot}=520^{+230}_{-140}\,R_\odot$
and an effective temperature of $T_\mathrm{eff}\,\sim\,2420\pm200$\,K.
The latter value is consistent with the effective temperature
of the used model M21n.
The combined model can reproduce the shape and
features of the observed photometry and visibility spectra of RR~Aql well,
as well as the SED at 1--40\,$\mu$m.



We conclude that our RR~Aql data can be described by a silicate dust shell 
surrounding a pulsating atmosphere, consistent with observations
of Mira variables \citep{Lorenz-Martins2000}.
The
effects of the pulsation on the mid-infrared flux and
visibility values are expected to be less than about
25\% and 20\%, respectively, and are too low to be detected
within our measurement uncertainties. Although the addition
of an Al$_2$O$_3$ dust shell did not improve the model fit,
our simulations also indicate that we cannot exclude the presence 
of an inner Al$_2$O$_3$ dust shell with relatively low optical depth,
which may be an important contributor to the dust condensation sequence. 

Our modeling attempt at a radiative transfer model of the dust
shell surrounding a dynamic dust-free model atmosphere 
provides constraints on the geometric extensions of atmospheric molecular
and dust shells. It should be noted that it cannot explain the 
mechanism by which the observed mass loss is produced and the wind
is driven. Newer models of M type, i.e. of oxygen-rich Miras,
that include atmospheric dust \citep[e.g. ][Ireland,
in preparation]{Ireland2008} typically predict effective acceleration very close to zero
in high layers but, so far, no clear outward acceleration, i.e. no wind.
Subtle details are being discussed that may overcome this shortcoming
of dynamic models of M type Miras \citep[e.g.][]{Hofner2011}.
Future observations aiming at characterizing and constraining other new  
models of the mass-loss process and the wind driving mechanism
at mid-infrared wavelengths would benefit from obtaining
more precise photometry values using a dedicated instrument
like VISIR at the VLT, the addition of shorter baselines
to characterize the extension of the silicate dust shell,
and a more complete coverage of the pulsation cycle. The addition
of concurrent, spectrally resolved, near-infrared interferometry
would be needed to more strongly constrain atmospheric molecular layers
located close to the photosphere.

\begin{acknowledgements}
We gratefully acknowledge star observers around the world whose 
devoted observations provide active data support in the observations 
of the variable stars. We used the AAVSO International Database 
along with the AFOEV and SIMBAD databases, operated at the CDS, France. 
The authors would also like to acknowledge all who are involved in 
developing the publicly accessible MIDI data reduction software packages 
and tools EWS, MIA, and MyMidiGui, namely Walter Jaffe, Rainer K{\"o}hler, 
Christian Hummel, and others. This work is based on service mode 
observations made with the MIDI instrument, which is operated by ESO. 
We would like to thank the operating team at the Paranal Observatory
for their careful execution of the observations.
\end{acknowledgements}

\bibliographystyle{aa}
\bibliography{rraql_karovicova}

\end{document}